\documentclass[twocolumn]{aastex631}
\usepackage{amsmath}
\usepackage{wrapfig}
\usepackage{svg}
\usepackage{comment}
\usepackage{booktabs}

\newcommand{\Mmw}{M_{\rm MW}}

\renewcommand{\l}{\left}
\renewcommand{\r}{\right}

\newcommand{\Tc}{T_{\rm col}}

\newcommand{\RB}{R_{\rm B}}

\newcommand{\lrp}[3]{\left(\frac{#1}{#2}\right)^{#3}}
\usepackage{comment}

\newcommand{\dd}{{\rm d}}
\newcommand{\pc}{~{\rm pc}}

\newcommand{\Gyr}{~{\rm Gyr}}

\newcommand{\Msun}{{M_{\odot}}}
\newcommand{\Rsun}{{R_{\odot}}}
\newcommand{\Rs}{R_\star}

\newcommand{\ns}{n_\star}
\newcommand{\bs}{b_\star}

\newcommand{\mb}{m_\bullet}

\newcommand{\nb}{n_\bullet}

\newcommand{\trg}{T_{\rm RG}}

\newcommand{\vesc}{v_{\rm esc}}

\newcommand{\kms}{~{\rm km/s}}

\newcommand{\sgr}{{\rm Sgr}~{\rm A}^*}

\graphicspath{{graphs/}}

\begin{document}

\title{Red Giant Destruction by Stellar and Black Hole Collisions in Galactic Nuclei}

\correspondingauthor{Barak Rom}
\email{barak.rom@princeton.edu}
\author[0000-0002-7420-3578]{Barak Rom}
\affiliation{Department of Astrophysical Sciences, Princeton University, Princeton, NJ 08544, USA}

\author[0000-0001-9185-5044]{Eliot Quataert}
\affiliation{Department of Astrophysical Sciences, Princeton University, Princeton, NJ 08544, USA}


\begin{abstract}
\noindent
We study the impact of collisions involving red giants (RGs) in the dense stellar environments of galactic nuclei. 
We analytically estimate when collisions with main-sequence stars or stellar-mass black holes can strip a RG's envelope via ram pressure or accretion-driven shocks, or eject its helium core through gravitational recoil.
At high velocities, $v\gtrsim10^3\kms$, collisions with main-sequence stars efficiently deplete the RG population. 
At lower velocities, collisions with stellar-mass BHs typically dominate over stellar encounters, but the overall RG destruction rate is low and does not significantly affect the RG population.
Nonetheless, these collisions produce low-mass helium white dwarfs, which are the stripped cores of the disrupted RGs, at a rate of $\sim 500\Gyr^{-1}$.  Helium white dwarfs can produce an interesting class of white dwarf tidal disruption events around $\sim 10^{5-6} M_\odot$ massive black holes where Carbon-Oxygen white dwarfs cannot be tidally disrupted outside the horizon.
Applied to our own Galactic Center, we quantify the impact of collisions on the observed population of RGs, as well as the effects of their intrinsic scarcity due to short RG lifetimes.   We find that the RG projected density flattens within~$\sim1$'', primarily due to collisions for fainter RGs and their short lifetimes for more luminous RGs.  
\end{abstract}


\section{Introduction}
Observations of the Galactic Center (GC) provide a unique opportunity to study the immediate environment of the nearest supermassive black hole (SMBH), located only $8~{\rm kpc}$ from earth, and the complex stellar dynamics surrounding it.

Over the past three decades, increasingly detailed observations have enabled accurate measurements of the SMBH mass, $\Mmw\simeq4.3\times10^6\Msun$ \citep{Ghez_2008,Gillessen_2009}, as well as tests of general relativity, including detections of orbital precession and gravitational redshift \citep{Grav_col_2018,Do_2019,Grav_col_2020,Grav_col_2022b}, as well as direct imaging of the SMBH's shadow \citep{EHT_SgrAI_2022}. 
These observations have also characterized the surrounding stellar population, revealing the puzzling presence of a cluster of young, massive stars within the central arcsecond \citep{Genzel_2003,GhezDuch_2003,GenEisGil_2010}, alongside a power-law cusp of older, fainter stars \citep{Schodel_2018}.

Focusing on the old stellar population, a power-law cusp is expected as the steady-state result of two-body scatterings \citep{Alexander_2017}, although the observationally inferred densities, $n\l(r\r)\propto r^{-\gamma}$ with $\gamma\approx1.1-1.4$ \citep{Cano_2018}, are somewhat flatter than the theoretically values of $\gamma\approx3/2-7/4$ \citep{BW_76,BW_77}. 
This may indicate the influence of additional dynamical processes that can flatten the stellar distribution, such as loss-cone depletion \citep{Cohn_1978,Bar_or_2016}, repeated episodes of star formation \citep{Aharon_2016,Baumgardt_2018}, or stellar collisions, as further discussed below. 

Luminous red giants (RGs), with K-band magnitudes $K\lesssim15-16$, appear to follow a power-law profile similar to that of the older stars at larger radii, yet their projected surface density flattens within the inner $\lesssim1-10$''.
This deficit, known as the ``missing red giants'' problem, was first identified in the 1990s \citep{Sellgren_1990,Genzel_1996} and has persisted since \citep[e.g.,][]{Eisenhauer_2005,Do_2009,Buchholz_2009,Bartko_2010,Do_2013}. 
However, more recent, higher-sensitivity imaging and spectroscopic studies suggest that the discrepancy is less severe than originally seemed \citep{Habibi_2019,Schodel_2020}. 

Several mechanisms have been proposed to explain this lack of RGs, including collisions with the ambient stellar population \citep{Genzel_1996,Alexander_1999, Bailey_1999, Dale_2009}, interaction with a clumpy gaseous disk \citep{Amaro_2014,Amaro_2020}, encounters with a jet during a past active galactic nuclei phase \citep{Zajaek_2020}, and tidal disruption by the SMBH, although the latter appears to be disfavored \citep[see][]{Kim_2026}.

More broadly, stellar collisions in nuclear stars clusters have been extensively studied, exploring the various possible collision outcomes between two MS stars \citep{Lai_1993,Rauch_1999,Freitag_2005,Rose_2026}, their impact on the stellar population properties and spatial distribution \citep{Murphy_1991,Freitag_2002,Rose_2023,Balberg_2024,Rose_2024,Gibson_2025,Rom_2026}, as well as their potential to produce observable  transients \citep{Balberg_2013,Amaro_2023,Ryu_2024}. 

In this work, we revisit the impact of RGs collisions in nuclear stellar clusters. 
In \S\ref{sec:Theo_frame} we present our analytical framework, including the RG stellar model and the stellar distributions in nuclear stellar cluster.
In \S\ref{sec:col}, we examine the conditions for destructive collisions between RGs and either MS stars or a stellar-mass black holes (BHs), and calculate their rates. 
In \S\ref{sec:WD} we estimate the abundance of He white dwarfs (WDs) produced as the stripped cores of the disrupted RGs.
In \S\ref{sec:miss_RG}, we apply our results to the ``missing red giants'' problem in the Galactic Center. We conclude in \S\ref{sec:sum}.

\section{Theoretical framework} \label{sec:Theo_frame}
We study the impact of collisions on RGs in dense nuclear stellar clusters.
We first describe the RG structure used to determine the collision outcomes, along with the adopted stellar distributions that enter the collision rate estimates.

\subsection{Red Giant structure \& evolution}
We model the structure and evolution of a solar-mass RG using the stellar evolution code MESA \citep{MESA_2011,Mesa_2013,MESA_2015,MESA_2018,MESA_2019,MESA_2023}.

The timescale spent at radii $\sim R$ along the RG branch can be estimated as $\trg\propto m_c/L\propto R^{-1.7}$, using that the luminosity scales as $L\propto\Rs^2$ and the core mass $m_c\propto L^{1/\eta}$, with $\eta\approx5-7$ \citep{Kippenhahn_2012}. 
A numerical fit to the MESA models yields a slightly shallower dependence
\begin{equation} \label{eq:Trg}
    \trg=\frac{R}{\dot{R}}\approx 1\Gyr \lrp{R}{2\Rsun}{-\xi},
\end{equation}
where $\xi\approx\left\{\def\arraystretch{1}\begin{tabular}{@{}l@{\quad}l@{}}
        $5$ & $R\lesssim 2\Rsun$ \\
        $1.4$ & $R\gtrsim 2\Rsun$\end{tabular}\right.$.
The critical radius $R\approx2\Rsun$ is roughly where the star develops a clearly differentiated helium core–envelope structure, which affects its evolution timescale and its sensitivity to collision, as discussed below.

From the same models, we obtain the envelope binding energy, 
\begin{equation}\label{eq:E_bin}
    E_{\rm bin}\approx3\times10^{47}\ {\rm erg} \lrp{R}{10\Rsun}{-0.8},
\end{equation}
{defined as the total energy required to unbind the stellar layers exterior to the He core}, and the core mass,
\begin{equation}
    m_c(R)\approx 0.25\Msun\left(\frac{R}{10\Rsun}\right)^{0.2}.
\end{equation}
These imply an escape velocity for the core to escape its surrounding envelope of
\begin{equation} \label{eq:vesc_core}
    \vesc=\sqrt{\frac{2E_{\rm bin}(R)}{m_c(R)}}\approx 350 {\rm km/s} \lrp{R}{10\Rsun}{-0.5}.
\end{equation}

\subsection{Nuclear cluster model}
We focus on the dynamics within the radius of influence of the SMBH, $R_h$, where it dominates the gravitational potential \citep{Merritt_2004}.
For the Milky-way galaxy, we adopt $R_h= 3 \pc$ \citep{Schodel_2018}.

For the density distributions of the stars and stellar-mass BHs, we consider a dynamically relaxed stellar cusp \citep{Alexander_2017}. 
Since collisions become frequent only in the densest inner regions of the nuclear cluster, we assume that stellar-mass BHs dominate the scattering\footnote{The BH cusp is expected to extend out to $\sim0.1R_h$, for a BH number fraction of $\sim10^{-3}$, and to decline steeply beyond this radius \citep[see discussion in][and references therein]{Rom_2025}.} and, hence, follow a Bahcall-Wolf cusp \citep{BW_76},
\begin{equation}\label{eq:nb}
    \nb(r)= \frac{M/\Msun}{4\pi R_h^3}\lrp{\mb}{\Msun}{-3/2}\left(\frac{r}{R_h}\right)^{-7/4},
\end{equation}  
where $\mb=10\Msun$ is the BH mass.

For the stars we adopt the following fiducial profile
\begin{equation}\label{eq:ns}
    \ns(r)= \frac{M/\Msun}{4\pi R_h^3}\lrp{r}{R_h}{-3/2},
\end{equation}
which arises in the region where stars are scattered by stellar-mass BHs \citep{BW_77}.
Nonetheless, in the following calculation we show the range of results corresponding to varying the stellar density power slope, between $\gamma=5/4$, as expected from the balance between stellar collisions and injection of stars via the Hills mechanism \citep{Rom_2026}, and $\gamma=7/4$, the single-mass Bahcall-Wolf cusp. 
The former, collision-regulated profile, better aligns with the observed density slope $\gamma\approx 1.1-1.4$, \citep{Cano_2018,Schodel_2018}.

In all cases, the normalization of the stellar density is fixed such that a Milky Way-like galaxy contains $\approx10^4$ stars within $0.1 \pc$, as inferred by observations of the GC \citep{Schodel_2018}.
For the BH density, Eq.~(\ref{eq:nb}), the normalization of $\left(\mb/\Msun\right)^{-3/2}\approx0.03$ relative to the stellar density (Eq.~\ref{eq:ns}) characterizes the region in which BHs take over as the dominant scatterers \citep{Linial_2022,Rom_2025}.

For the RGs, we assume that in regions where collisions are negligible, the RG density profile follows that of their MS progenitors,
\begin{equation}\label{eq:nrg}
\begin{aligned}
    n_{\rm RG}\left(r,R\right)&= A_0\frac{\trg(R)}{T_{\rm MS}\left(\Msun\right)}n_{\star}(r),
\end{aligned}
\end{equation}
 where the normalization is set by the ratio of the RG timescale (Eq.~\ref{eq:Trg}) to the MS lifespan, $T_{\rm MS}\left(\Msun\right)\approx10\Gyr$.
The prefactor $A_0\simeq0.7$ is calibrated to match the observed density of bright RGs\footnote{RGs with K-band magnitude $K<16$, corresponding roughly to $R\gtrsim10\Rsun$. We relate the K-band luminosity to the RG radius using absolute magnitudes computed with MESA and an extinction parameter $A_k=2.62$ \citep{Schodel_2020}.\label{fn:k_band} } at a distance of $1\pc$ from $\sgr$ \citep{Cano_2018}. 

In this case, the RGs inherit the isotropic distribution of the MS stars. Therefore, at a fixed semimajor axis $\sim a$, the fraction of RGs with pericenters smaller than a given value $r_p$ scales linearly with $r_p$.

The RG average occupation number falls below unity in the orbital phase-space, i.e., in the $\l(a,r_p\r)$ plane, beneath the line given by
\begin{equation} \label{eq:rpN1}
\begin{aligned}
    \frac{r_p}{R_h}\approx&2\times10^{-6}\lrp{a}{R_h}{-1/2}\lrp{R}{\Rsun}{\xi}\\
    &\times\lrp{M}{\Mmw}{-1}.
\end{aligned}
\end{equation}
This corresponds to a characteristic radius for circular orbits
\begin{equation} \label{eq:aN1}
\begin{aligned}
    \l.\frac{a}{R_h}\r|_{\rm N=1}\approx&10^{-3}\lrp{R}{10\Rsun}{0.93}\lrp{M}{\Mmw}{-2/3}.
\end{aligned}
\end{equation}

Another outcome of the RGs' short lifetime is that two-body scattering may significantly alter their orbit only at a restricted region in the $\l(a,r_p\r)$ plane, below the line
\begin{equation} \label{eq:rp_2b}
    r_{p,\rm 2B} \l(a,R\r) \approx 0.25\lrp{a}{R_h}{3/4}\lrp{R}{\Rsun}{-1.4},
\end{equation}
given by equating the RG lifetime (Eq. \ref{eq:Trg}) and the two-body scattering timescale\footnote{In the two-body scattering timescale, Eq. (\ref{eq:T2b}), we assume that the scatterers follow a Bahcall-Wolf cusp \citep{Linial_2022,Rom_2025}.} \citep{Biney_Tremaine}
\begin{equation}\label{eq:T2b}
    T_{\rm 2B}\left(a,r_p\right)\approx 8\Gyr \lrp{M}{\Mmw}{5/4}\lrp{a}{R_h}{1/4}\lrp{a}{r_p}{}.
\end{equation}

Above the line defined in Eq.~(\ref{eq:rp_2b}), the time it takes for two-body scattering to change the orbit exceeds the RG lifetime. Therefore, in this region, stellar evolution proceeds effectively at a fixed orbit. Notably, for low-eccentricity orbits, the scattering timescale can exceed the entire RG phase.

Finally, in the context of tidal disruption events \citep[TDEs;][]{Rees_88}, RGs can reach their tidal radius either through scattering onto highly eccentric orbits - analogous to the standard MS-star TDEs \citep{Magorrian_1999}, though here limited to the smaller scattering-dominated region - or by evolving from MS stars that are already on orbits within the RG loss cone \citep{Syer_1999,MacLeod_2013}, as the RG tidal radius is roughly $R_{t}/R_{t,\odot} \approx R/\Rsun$, where $R_{t,\odot}\approx 20R_{g}$ is the tidal radius of a solar-mass MS star and $R_g=G\Mmw/c^2\approx2\times10^{-7}\pc$ is the SMBH gravitational radius.

 \begin{figure*}[ht!]
    \centering
    \includegraphics[width=\linewidth]{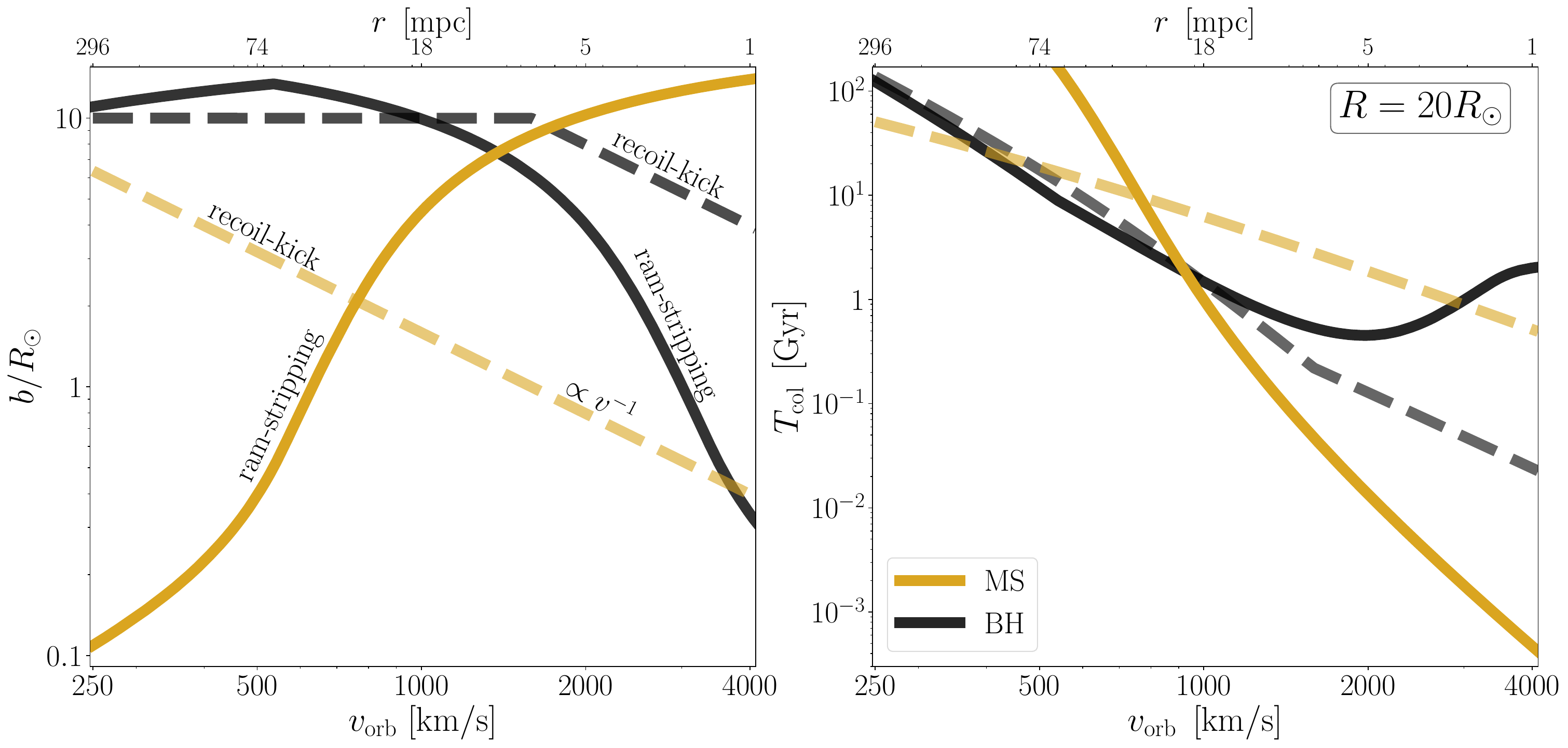}
    \caption{Critical impact parameter for destructive collisions (left panel) and the corresponding collision timescale (right panel) as a function of orbital velocity for a $R=20\Rsun$ RG.
    We consider both stripping of the RG envelope via ram-pressure-driven shocks (solid lines) and ejection of the He core by gravitational recoil (dashed lines). 
    In each case, the impactor is either a solar-mass MS star (yellow lines) or a $10\Msun$ stellar-mass BH (black lines). 
    The top x-axis shows the distance from the SMBH for a Milky Way-like galaxy.}
    \label{fig:impact_parameter}
\end{figure*}
\section{Red giant collisions} \label{sec:col}
We examine two types of collision outcomes that can disrupt the RGs: stripping the envelope via ram-pressure-generated shock (\S\ref{sec:ram}) and ejecting the He core by a gravitational recoil kick (\S\ref{sec:kick}).

Generally, the characteristic collision timescale is given by
\begin{equation} \label{eq:Tc}
    \Tc\approx\left[n_i\sigma_i v\right]^{-1},
\end{equation}
where $i\in\left(\star,\bullet\right)$ denotes the impactor, MS star or stellar-mass BH, $n_i$ is the corresponding impactors number density, $\sigma_i$ is the cross section, and $v$ is the relative velocity, which we approximate as the local orbital velocity.

For each collision type, we identify the critical impact parameter $b_i$, corresponding to the maximal distance for which the encounter can disrupt the RG. 
This then sets the effective cross-section for the collision \citep{Biney_Tremaine}
\begin{equation}\label{eq:Sig}
    \sigma_i\approx\pi b_i^2\left(1+\frac{2G\left(m_i+\Msun\right)}{b_iv^2}\right),
\end{equation}
where $m_i$ is the impactor mass. 

The probability that an RG with radius of order $\sim R$ undergoes a destructive collision is then
\begin{equation}\label{eq:p_i}
\begin{aligned}
    p(v,R)&=\frac{\trg(R)}{\Tc(v,R)},
\end{aligned}
\end{equation}
where $\trg\left(R\right)$, as given in Eq.~(\ref{eq:Trg}), is the time an RG has a radius of order $R$ along its evolution. 

Consequently, the survival probability, i.e., the probability that an RG evolves up to radius $R$ without being destroyed by collisions, is
\begin{equation}\label{eq:P}
    \mathcal{P}\left(v,R\right)=\exp\left[-\int_{\Rsun}^R p(v,R')\frac{dR'}{R'}\right],
\end{equation}
\begin{figure*}[ht!]
    \centering
    \includegraphics[width=\linewidth]{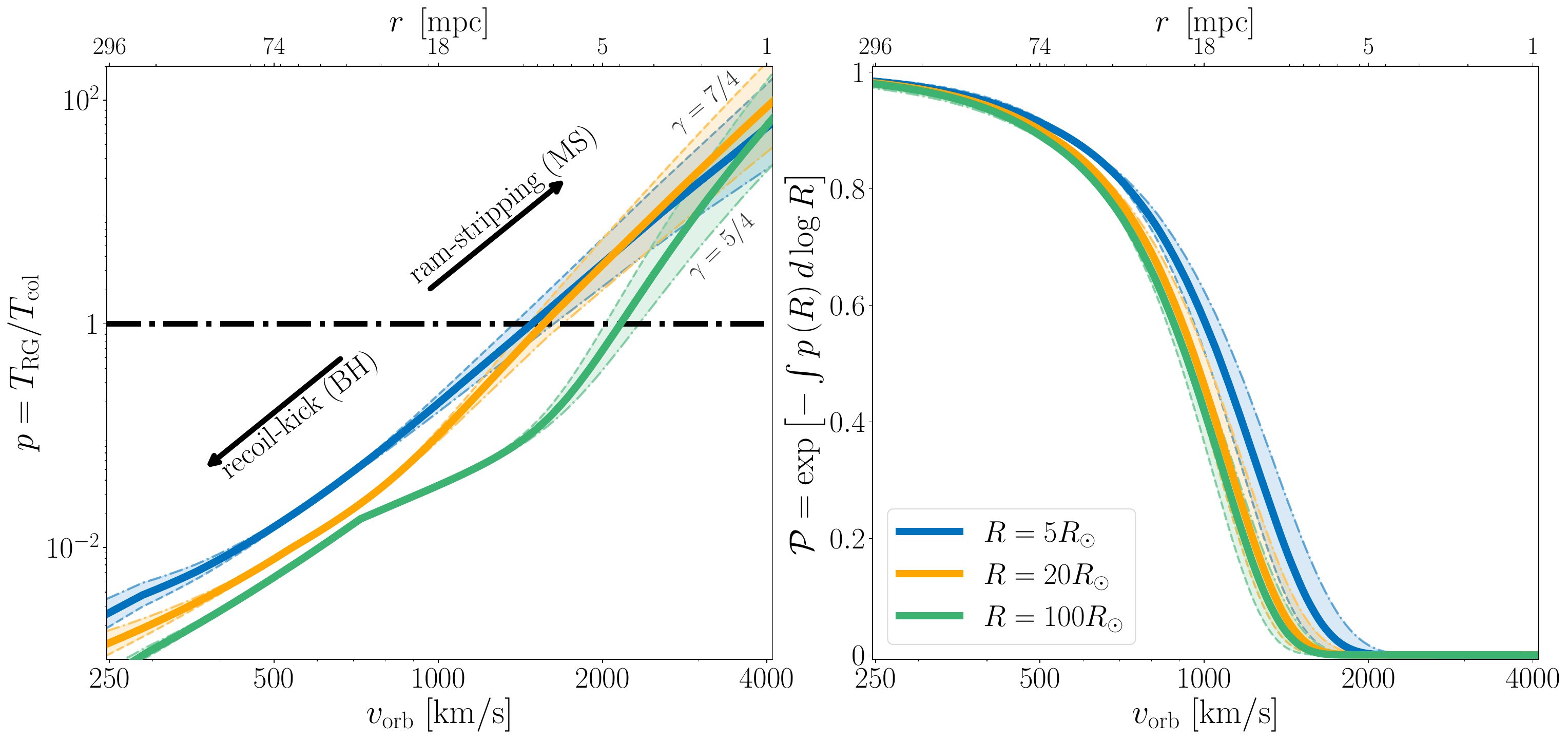}
    \caption{Collision probability as a function of orbital velocity for RG radii $R=5\Rsun$ (blue), $R=20\Rsun$ (yellow), and $R=100\Rsun$ (green) in a Milky Way-like galaxy. 
    The left panel presents the probability an RG with a given stellar radius $R$ would undergo a destructive collision (Eq. \ref{eq:p_i}), while the right panel presents the integrated survival probability (Eq. \ref{eq:P}).
    At low velocities, $v\lesssim500\kms$, the rate of RG  collisions is negligible and it is dominated by encounters with stellar-mass BHs. At high velocities, $v\gtrsim1500\kms$, stripping of the RG envelope by MS stars becomes the dominant channel, efficiently depleting the RG population.
    The shaded regions illustrate the impact of varying the density slope for the background stellar distribution, from the fiducial $\gamma=3/2$ to $\gamma=5/4$ (dashed-dotted boundary) and $\gamma=7/4$ (dashed boundary)  
    The top x-axis shows the distance from the SMBH.}   
    \label{fig:coll_prob}
\end{figure*}

\subsection{Ram-pressure envelope stripping} \label{sec:ram}
At velocities exceeding the envelope's sound speed, $c_s\sim100\kms$, an impactor passing through the RG can generate a shock that strips its envelope if the energy deposited $E_d$ is comparable to the envelope's binding energy
\begin{equation}\label{eq:E_ram}
    E_d \simeq \Sigma_{\rm den}(b,R)v^2\pi \widetilde{R}_{\rm i}^2\gtrsim E_{\rm bin}(R),
\end{equation}
where $\Sigma_{\rm den}\left(b,R\right)$ is the column mass density of an RG of radius $R$ along a trajectory with impact parameter $b$.
The effective impactor radius is defined as $\widetilde{R}_{\rm i}=\max\left\{R_{i},\RB\right\}$, where $R_{i}$ is the physical radius of the impactor and $\RB=Gm_i/\l(v^2+c_s^2\r)$ is the Bondi-Hoyle radius, evaluated at a radial distance $b$ from the center of the RG.

Since the pressure (and the sound speed) weakly varies throughout most of the RG envelope, as further discussed in \S\ref{sec:P_vs_E}, a highly supersonic impactor drives a strong shock that penetrates deep into the envelope rather than remaining locally confined, and can therefore unbind significant fraction of the envelope.   

In Eq.~(\ref{eq:E_ram}), we assume that the ram pressure of the impactor is fully converted into shock energy that penetrates deeper into the star. Realistically, this conversion should include an efficiency prefactor. However, because the collision timescale to destroy RGs depends very steeply on velocity (see Eq.~\ref{eq:ps} and Fig.~\ref{fig:coll_prob}), even adding a dimensionless prefactor $\sim\mathcal{O}(0.1)$ in Eq.~(\ref{eq:E_ram}) only produces an order-unity shift in the velocity where collisions become important (i.e., where $p\approx1$), since a small increase in velocity offsets the reduced efficiency.

Furthermore, in the following calculation, we consider encounters that strip the RG envelope in a single collision. 
We do not account for repeated collisions since they have a negligible effect on the overall RG depletion rate, as shown in Appendix~\ref{app:b}.

Figure~(\ref{fig:impact_parameter}) presents the critical impact parameter, defined as the maximal $b$ that satisfies Eq.~(\ref{eq:E_ram}), for MS star (yellow solid line) and stellar-mass BH (black solid line) impactors.    At high velocities, $v\gtrsim 10^3\kms$, MS-RG collisions dominate, efficiently depleting the RG population.
At somewhat lower velocities, $500\kms\lesssim v\lesssim 10^3\kms$, the impact parameter for MS-RG ram-stripping collisions strongly depends on the velocity, increasing from $\bs\ll\Rsun$ to $\bs\sim R/2$.

This significant increase of the impact parameter in the intermediate velocity range can be understood by an order of magnitude estimate.
We approximate the envelope density as a power law
\begin{equation} \label{eq:rho_app}
\rho(r)\sim \frac{\Msun}{R^{3}}\lrp{r}{R}{-1.5},
\end{equation}
which is a reasonable approximation for the bulk of the envelope, roughly from $\sim0.5\Rsun$ to $\sim 0.5 R$, away from both the He core and H-burning shell as well as the outer layers where the density rapidly drops to zero. 
The corresponding column mass density can be estimated as
\begin{equation} \label{eq:sig_app}
\Sigma_{\rm den}\left(\bs,R\right)\sim  \bs\rho\left(\bs\right) \sim \frac{\Msun}{R^2}\lrp{\bs}{R}{-0.5}.    
\end{equation}

Thus, from Eqs. (\ref{eq:E_ram}) and (\ref{eq:sig_app}), the impact parameter scales as $\bs \propto v^{4}R^{-1.3}$, while the collision timescale follows $\Tc \propto v^{-12}R^{2.6}$. 
Consequently, the collision probability scales as 
\begin{equation} \label{eq:ps}
    p\propto v^{12}R^{-4}.
\end{equation} 

A modest increase in velocity therefore leads to a sharp rise in the  collision probability. 
A more detailed derivation of the impact parameter under the power-law density approximation is given in Appendix~\ref{app:a}.

The numerical results indeed show a steep dependence of the collision probability on velocity (see Fig.~\ref{fig:coll_prob}), though somewhat shallower than that predicted by the simple order-of-magnitude estimate (Eq.~\ref{eq:ps}). 
This discrepancy is expected, as the power-law density profile approximation breaks down at both small and large impact parameters, which are approached in the low- and high-velocity limits, respectively. 
For larger RGs, however, this approximation holds over a broader range of impact parameters and velocities. 

At low relative velocities, $v\lesssim500\kms$, stellar-mass BHs typically dominate the RGs collision rate, especially for larger RGs, and core ejection via gravitational recoil becomes a significant disruption channel (as discussed in \S\ref{sec:kick}). 

Notably, the Bondi-Hoyle radius can become comparable to the RG radius at velocities of order a few hundred km/s. 
Consequently, the interactions become global and the impulsive-collision approximation, underlying our estimate of the deposited energy in Eq.~(\ref{eq:E_ram}), breaks down.
A full treatment of collisions in this regime is beyond the scope of this work. 
However, such low-velocity encounters are rare and have a negligible impact on the RG population.
We therefore adopt a simple cutoff for the effective impactor radius, requiring $\widetilde{R}_{\rm i} \lesssim (R - b)$.

\subsubsection{Can the collision strip the envelope?} \label{sec:P_vs_E}
In addition to the energy requirement, namely, that the deposited energy exceeds the envelope binding energy (Eq.~\ref{eq:E_ram}), shock-driven mass loss can only expel material exterior to the layer where the effective ram pressure becomes comparable to the local stellar pressure. A similar argument has been used to estimate mass loss in the context of star-disk collisions \citep[e.g.,][]{LuQua_2023,Linial_Metzger_2023, Yao_2025b}, and interaction of supernovae ejecta with a companion star \citep[e.g.,][]{Marietta_2000,Hirai_2018,Wong_2025}.  

One difference here is that the shock/drag energy is initially imparted to a small volume of the star (the column interacted with by the impactor). 
Because the work done by the drag force is dominated by $r \sim b$, we estimate the effective pressure of the shock this drives by assuming that the  energy deposited (eq. \ref{eq:E_ram}) is distributed throughout a volume $\sim b^3$.

Accordingly, we require
\begin{equation} \label{eq:P_eff} 
    P_{\rm eff} \equiv \frac{E_d}{4/3 \pi b^3} \approx P_{\rm ram}\lrp{\widetilde{R
    }_{\rm i}}{b}{2}\gtrsim P^{*},
\end{equation}
where $P_{\rm ram}\approx \rho\l(b\r)v^2$, so the shock remains supersonic down to the depth where the stellar pressure reaches $P^{*}$. 
As a fiducial value, we take $P^{*}$ to be the pressure at the radial shell enclosing $0.45\Msun$. 
This choice reflects the approximate maximal core mass prior to helium ignition; if the remaining bound mass falls below this threshold, we expect that the subsequent RG evolution would be significantly altered.

\begin{figure}[ht!]
    \centering
    \includegraphics[width=8.6cm]{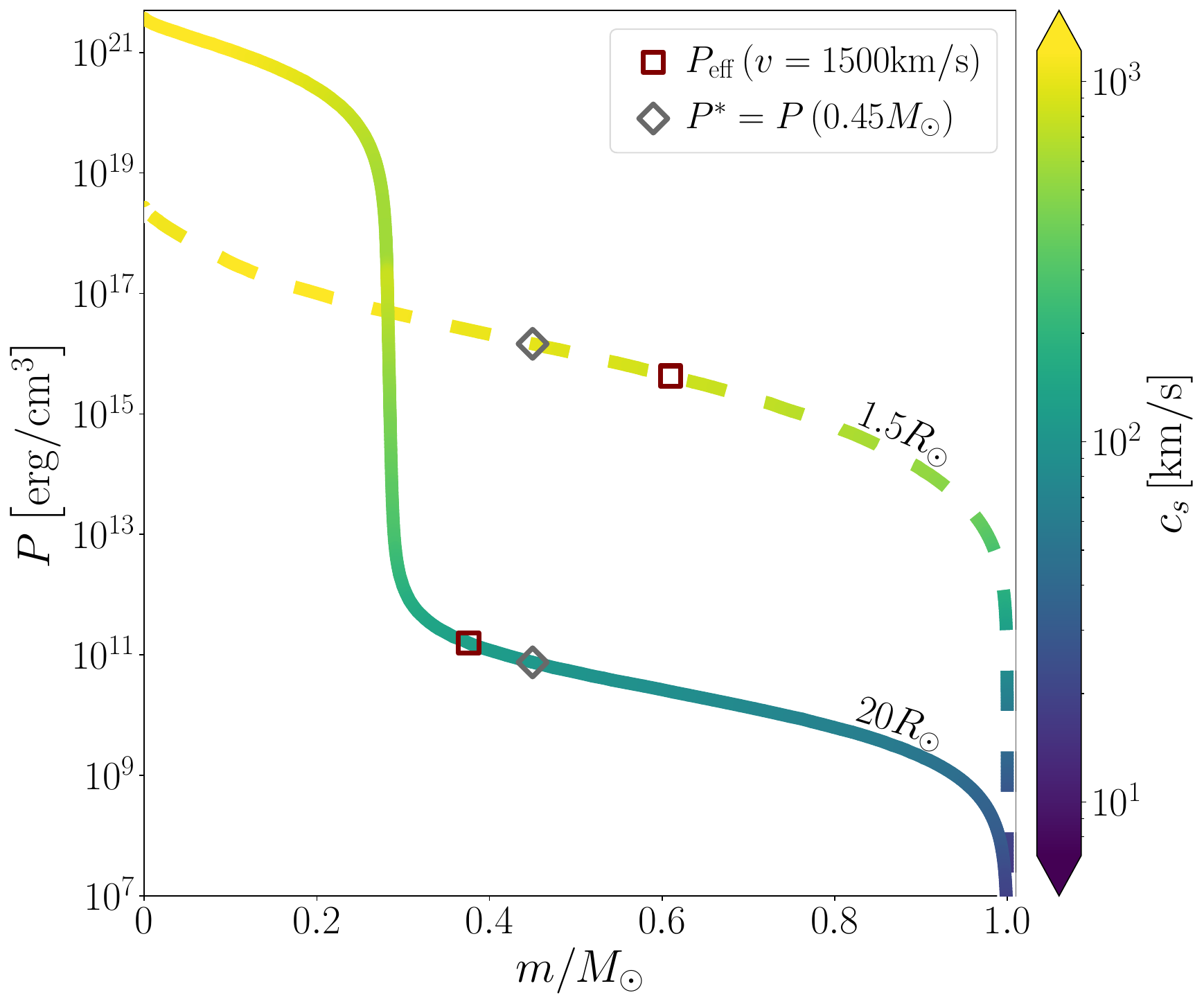}
    \caption{RG pressure profiles as a function of enclosed mass for $R=1.5\Rsun$ (dashed line) and $R=20\Rsun$ (solid line). The colormap shows the local sound speed. The gray diamonds mark $P^*$, the pressure at the shell enclosing $0.45\Msun$. 
    The red squares show the effective ram pressure generated during a collision with an MS star at $v=1500\kms$ and at the maximal impact parameter for which the deposited energy equals the envelope binding energy (Eq.~\ref{eq:E_ram}). 
    For the $20\Rsun$ RG, this effective ram pressure exceeds $P^{*}$, allowing the shock to penetrate deeply. In contrast, for the $1.5\Rsun$ case, the shock becomes subsonic around $m\approx0.5\Msun$. 
    The relative positions of the markers illustrate how envelope stripping may be limited by sufficient energy deposition (as in the RG case) or by the stellar pressure gradient (as for MS stars and subgiants).
    }
    \label{fig:pressure}
\end{figure}

Once the star develops a core-envelope structure, achieved for $R\gtrsim2\Rsun$,
the pressure varies weakly throughout most of the RG envelope, as shown in Fig.~(\ref{fig:pressure}). 
Therefore, a highly supersonic passage of an impactor can drive a strong shock that unbinds a substantial fraction of the envelope. 

To estimate whether Eq.~(\ref{eq:P_eff}) is indeed satisfied, we recast Eq.~(\ref{eq:E_ram}) as $P_{\rm eff}\gtrsim 0.25 \bar{P}\l(R/b\r)^3$, where we assume $\widetilde{R}_{\rm i}\ll b\lesssim R$ and denote $\bar{P}\approx G\Msun^2/R^4$. 
Thus, depositing the required amount of energy demands an effective ram pressure exceeding the average pressure $\bar{P}$ by a factor of $\sim\left(R/b\right)^3$. 
For RGs, $\bar{P}\sim P^{*}$ and $b\sim R/2$, implying that satisfying the energy requirement in equation \ref{eq:E_ram} necessarily leads to a shock pressure large enough to reach depths $\sim P^*$, leading to a substantial mass loss.   This is shown explicitly in Figure \ref{fig:pressure} by highlighting the relative locations where $P = P_{\rm eff}$ and $P = \bar P$.

By contrast, during the MS phase and the transition to the RG branch (``subgiants'', with $R\lesssim2\Rsun$), the pressure gradient within the star becomes the dominant constraint limiting how much mass is lost (see the sub-giant model in Fig. \ref{fig:pressure}).  Even if the relative kinetic energy of two MS stars exceeds the stellar binding energy, an even higher velocity is required for the shock to remain strong deep in the stellar interior and thus to unbind the stars during a collision.  
\begin{figure*}[ht!]
    \centering
    \includegraphics[width=\linewidth]{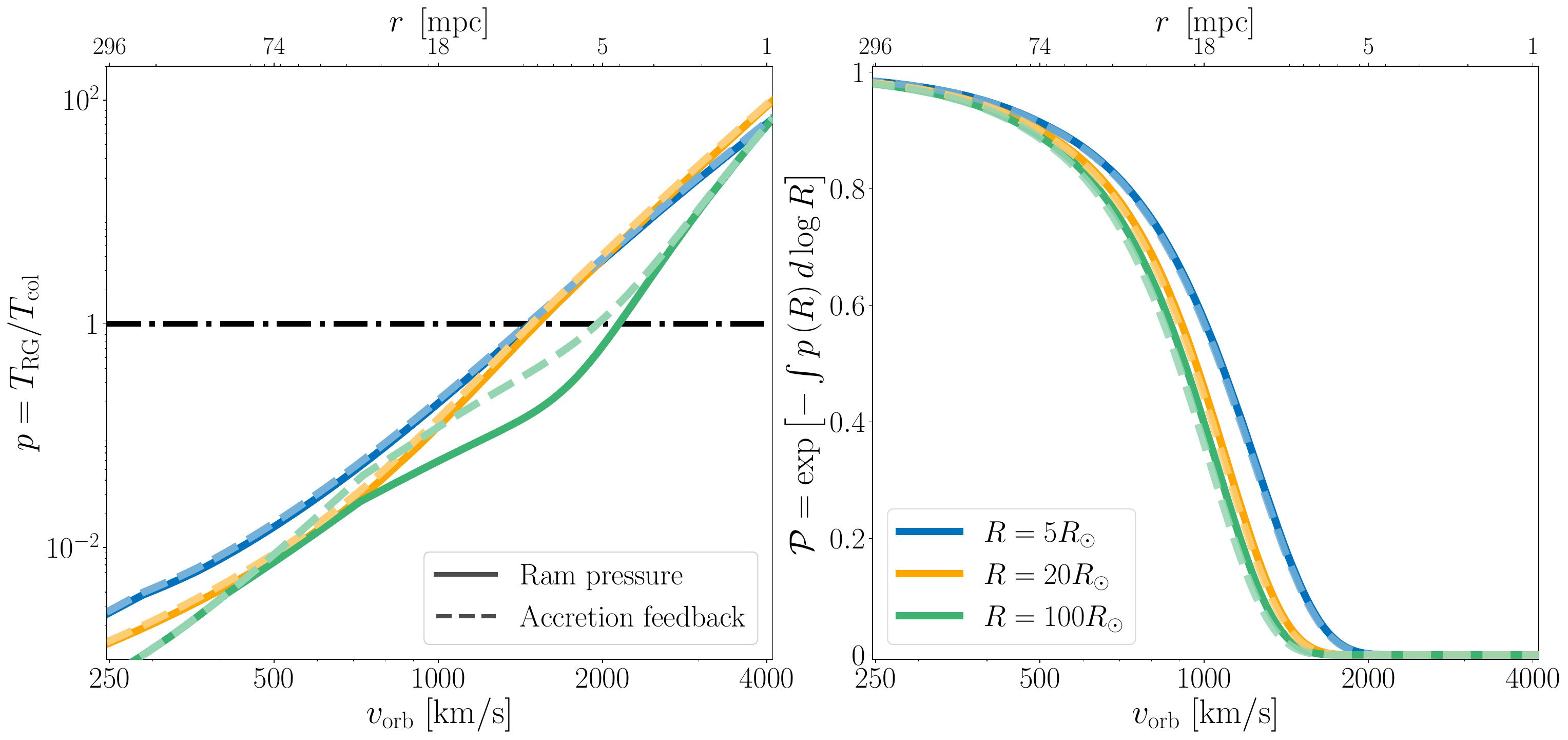}
    \caption{Same as Fig.~\ref{fig:coll_prob}, but including the effect of accretion feedback during encounters with stellar-mass BHs (dashed lines). 
    The feedback-enhanced stripping impact parameter increases the collision probability for larger RGs. Nonetheless, the survival probability remains only modestly affected, as the RG depletion is primarily set by the efficient destruction of smaller RGs at velocities $v\gtrsim10^3\kms$. For clarity, we show only the results for our fiducial slope $\gamma=3/2$ for the background stellar profile.}
    \label{fig:coll_prob_acc}
\end{figure*}
This is because Sun-like stars are highly centrally concentrated:  the central stellar pressure $P^*=P_c \approx 2\times10^{17}~{\rm erg/cm^{3}}$ satisfies $P^{*}/\bar{P}\approx 20$.  In addition, in the case of head-on collisions of two Sun-like stars, $P_{\rm eff}\approx P_{\rm ram}$. 
Consequently, the velocity required for a fully destructive, head-on collision is $v\approx \l(P_c/\bar{P}\r)^{1/2}v_{\rm esc,\odot}\approx 3\times10^3\kms$. 

Therefore, while a simple energy argument suggests that a velocity comparable to $v_{\rm esc,\odot}$ leads to a full disruption of the star, the contrast between the mean and central pressures implies a required velocity higher by a factor of $\approx4.6$. 
This estimate is simplistic in that it treats the shock as effectively planar neglecting geometric focusing.   
Our estimate is, however, consistent with the detailed numerical study of \cite{Rose_2026}. 
Expanding their fitting formula for the mass-loss fraction (Eq.~2) in the limit of a head-on, fully destructive collision yields $v \simeq 4.4 \, v_{\rm esc}$.

\subsubsection{Black hole accretion feedback}\label{sec:acc}
In the above calculation we estimated the drag force the impactor experience based on the ram pressure. 
However, a stellar-mass BH impactor may accrete from the envelope material as it passes through it. The exact accretion rate and the response of the envelope is uncertain, yet it might result in a significant accretion feedback, increasing the impact of the stellar-mass BH on the RG envelope.  We assume, based on recent GRMHD simulations of non-radiative flows \citep{Cho_2024,Guo_2025}, that the accretion rate onto the BH is given by
\begin{equation}
    \dot{m}_{\bullet} \sim \dot{m}_{\rm B}\lrp{R_g}{R_{\rm B}}{1/2},
\end{equation}
where $\dot{m}_{\rm B}\sim \rho\pi\RB^2\l(v^2+c_s^2\r)^{1/2}$ is the Bondi-Hoyle accretion rate and $R_g=G\mb/c^2\approx15~{\rm km}$ is the stellar-mass BH gravitational radius.

Taking $\dot{E}_{\rm acc}\sim \eta \dot{m}_{\bullet} c^2$, with $\eta\sim0.1$, and integrating it along the envelope crossing time, we get that the energy deposited by the accreting stellar-mass BH is
\begin{equation}\label{eq:E_acc}
    E_{\rm acc}\sim \eta\Sigma_{\rm den}(b,R)\pi \RB^2v^2\lrp{c}{v}{}\approx \eta \l(\frac{c}{v}\r)E_d.
\end{equation}

Consequently, the critical impact parameter for the stellar-mass BH to strip the envelope increases relative to the estimate derived in \S\ref{sec:ram}. 
This effect is most pronounced for larger RGs, for which the substantially larger impact parameter leads to a shorter collision timescale, as illustrated in Fig.~(\ref{fig:coll_prob_acc}).

Nonetheless, since at high velocities the depletion of RGs is set by the destruction of the smaller RGs, the impact on the overall survival probability, and therefore on the projected density, remains modest.

\subsection{Core gravitational recoil}\label{sec:kick}
In parallel with ram-pressure envelope stripping, the He core can be ejected by a gravitational recoil kick from a passing impactor \citep{Tuchman_1985,Dale_2009}. 

The core can be ejected from its surrounding envelope if the recoil velocity exceeds the escape velocity
\begin{equation}
    \Delta v\approx \frac{2Gm_i}{b v}\approx \vesc\l(R\r).
\end{equation}

Thus, we get
\begin{equation}\label{eq:b_kick}
b_i\approx\l(\frac{R}{2}\r)\times\min\l\{1,\frac{m_i}{2\Msun}\frac{\vesc\l(R\r)}{v}\r\}.
\end{equation}
At high velocities, $v/\vesc\l(R\r)\gtrsim0.5\l(m_i/\Msun\r)$, the critical impact parameter scales as $b_i\propto R^{0.5}v^{-1}$. 
At lower velocities, we limit the impact parameter to $b_i= R/2$, so the encounter can separate the core from the envelope rather than perturbing the RG as a whole.
In the left panel of Fig.~(\ref{fig:impact_parameter}), the dashed lines indicate the critical impact parameter for this type of encounters.

Ejection of the He core, especially from large RGs by stellar-mass BHs, contribute to the collision probability mostly at intermediate velocities, $v \lesssim 10^3\kms$; yet, collisions remain rare in this regime, as shown in Fig.~(\ref{fig:coll_prob}). 
Thus, overall, core ejection via gravitational recoil has only a marginal impact on the RG population and does not significantly alter the depletion due to envelope stripping by MS stars at high velocities.

Nonetheless, encounters between stellar-mass BHs and stars in the nuclear stellar clusters, either RGs (as discussed here) or MS stars, can lead to accretion onto the BH, producing luminous high-energy transients and providing a potential growth channel for stellar-mass BHs \citep[e.g.,][]{Kremer_2022,Rose_2022,Rizzuto_2022}
\begin{figure*}[ht!]
    \centering
    \includegraphics[width=\linewidth]{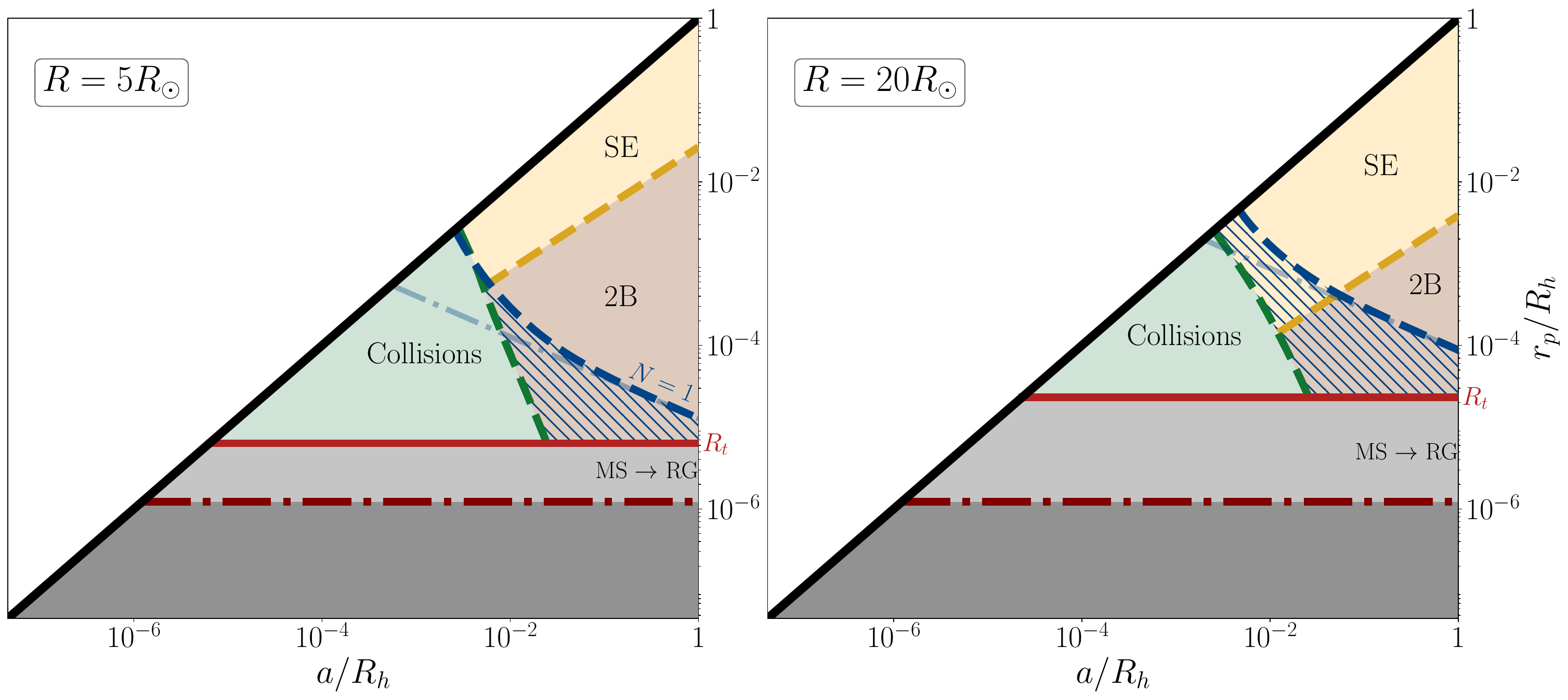}
    \caption{Orbital phase space in the semimajor axis–pericenter $(a,r_p)$ plane for RGs with radii $R=5\Rsun$ (left panel) and $R=20\Rsun$ (right panel), in a Milky Way–like galaxy.
    The phase space is divided into three regions: (a) stellar evolution–dominated (yellow), where RGs evolve without colliding or significant orbital scattering; (b) two-body scattering–dominated (brown); and (c) collision–dominated (green), where RGs of a given stellar radius are expected to undergo a destructive collision during their lifetime.
    The blue dashed line marks the $N=1$ contour, where the RG occupation number is unity. Collisions shift this contour to larger distances relative to the expectation from an isotropic distribution, shown by the faded blue dash–dotted line.
    The blue hatches highlights a region expected to be largely vacant of RGs, extending beyond the collision-dominated region.   
    The red horizontal lines indicate the tidal radii of an RG (solid) and an MS star (dash-dotted). The grey band between them denotes orbits where MS stars can reside and evolve into RGs, which are subsequently tidally disrupted by the SMBH.}
    \label{fig:tri}
\end{figure*}

\section{Formation rate of He White Dwarfs}\label{sec:WD} 
The destruction of RGs leaves behind low-mass He WDs, with masses below $\lesssim0.45\Msun$, which in the field are expected to form only in tight binaries \citep{ELM_Brown_2020}.
The He WD formation rate therefore directly traces the RG destruction rate
\begin{equation}\label{eq:rate_hewd}
    \begin{aligned}
    \frac{{\rm d}\Gamma_{\rm He-WD}}{{\rm d}\log \l(r\r)}&\approx\int {\rm d}\log\l(R\r)\frac{N_{\rm RG}\l(r,R\r)}{\Tc\l(r,R\r)}\\
    &=\frac{N_\star\l(r\r)}{T_{\rm MS}}\Delta\mathcal{P}\l(r\r),
\end{aligned}
\end{equation}
where $N_\star\l(r\r)$ is the number of stars per logarithmic radial bin and $N_\star\l(r\r)/T_{\rm MS}$ is the naive WD formation rate in the absence of collisions. 
The quantity $\Delta\mathcal{P}\l(r\r)=\mathcal{P}\l(r,R
_{\min}\r)-\mathcal{P}\l(r,R_{\max}\r)$ measures the fraction of RGs disrupted at a given distance from the SMBH. Here, $R_{\min}\approx2\Rsun$ corresponds to the stellar radii where a distinct core-envelope structure develops, and $R_{\max}\approx150\Rsun$ denotes the tip of the RG branch.

Integrating numerically over $r$ yields a total He WD formation rate of order
\begin{equation}\label{eq:rate_hewd_int}
    \Gamma_{\rm He-WD}\sim 500\Gyr^{-1},
\end{equation}
with a weak dependence on the SMBH mass.
The rate is dominated by the outer regions of the nuclear cluster. 
Although collisions are rare in these relatively low-velocity environments, and thus do not affect the overall RG population, this is offset by the larger number of RGs at large radii.

This can be understood from simple scalings. 
The number of RGs follows the stellar distribution\footnote{We obtain the scaling with the SMBH mass $M$, by adopting an $M\propto\sigma_h^4$ relation, where $\sigma_h$ is the stellar velocity dispersion \citep{Kormendy_2013}.} $N_{\rm RG}\propto M^{1/4}r^{3/2}$, while, in the limit $p\ll1$,  the disruption fraction scales as $\Delta\mathcal{P}\l(r\r)\approx\int p(r){\rm d}\log R \propto r^{-1}M^{-1/4}$. 
Combining these scalings implies that the WD formation rate is weighted toward larger radii and weakly depends on the SMBH mass.

The latter estimate, $\Delta\mathcal{P}\propto r^{-1}M^{-1/4}$, assumes that collisions are dominated by gravitational recoil, such that $b\propto v^{-1}$ (see Eq. \ref{eq:b_kick}), and neglects gravitational focusing. 
Numerically, we find $\Delta\mathcal{P}\propto r^{-1.16}$ for $r\geq0.1\pc$.

The He WD formation rate, Eq.~(\ref{eq:rate_hewd_int}), is approximately two to three orders of magnitude lower than the expected formation rate of C/O WDs, with typical masses $\approx 0.6\Msun$, which are the standard end products of solar-mass stellar evolution. 
Therefore, their formation rate can be estimated as
$\Gamma_{\rm C/O-WD}\sim N(R_h)/T_{\rm MS}\sim 10^5\l(M/\Mmw\r)^{1/4}\Gyr^{-1}$.

Although the He WDs represent only a small fraction of the WD population, they may produce unique transients.
In particular, while C/O WDs can be tidally disrupted only by intermediate mass BHs \citep[and are the primary focus in the literature; see][and references therein]{Maguire_2020}, with $M\lesssim10^5\Msun$, the lighter He WDs have larger tidal radii, especially if they retain a residual, shell-burning hydrogen envelope.
As a result, they can be disrupted by low-mass SMBHs $M\sim10^{5-6}\Msun$ \citep{Smith_2017}. 
We defer a detailed study of such transients to a subsequent paper.

\section{The ``missing red giants''}\label{sec:miss_RG}
Our results suggest that RGs are significantly depleted from the innermost regions of the nuclear stellar cluster due to destructive collisions. The RG survival probability (Eq. \ref{eq:P}) steeply dropping from unity to zero around velocities of order $\sim10^3\kms$, primarily driven by envelope-stripping MS-RG collisions.

Figure (\ref{fig:tri}) presents the resulting orbital phase space in the $(a,r_p)$ plane for $R=5\Rsun$ and $R=20\Rsun$ RGs in a Milky Way-like galaxy. 
In the innermost green collision-dominated region, RGs are expected to undergo destructive collisions within their lifetime.
The yellow stellar evolution-dominated region corresponds to orbits where the RG lifetime is shorter than both collision and two-body scattering timescales, hence, they evolve along the RG branch on roughly fixed orbits. 
In the brown scattering-dominated region, RGs can be significantly scattered during their lifetime. 

The boundaries between these regions are defined by equating the RG lifetime to the collision timescale (green dashed line) and to the scattering timescale (yellow dashed line; Eq. \ref{eq:rp_2b}).
The blue dashed line marks the $N_{\rm RG}=1$ contour, below which the RG occupation number falls below unity, accounting both depletion by collisions and the intrinsic RG abundance. 
For reference, the shaded dash-dotted blue line shows the same contour in the absence of collisions.
As pointed out by \cite{Kim_2026}, the depletion of RGs on highly eccentric orbits, as evident in Fig. (\ref{fig:tri}), implies an anisotropy that may be observationally detectable.

\begin{figure}[ht!]
    \centering
    \includegraphics[width=8.6cm]{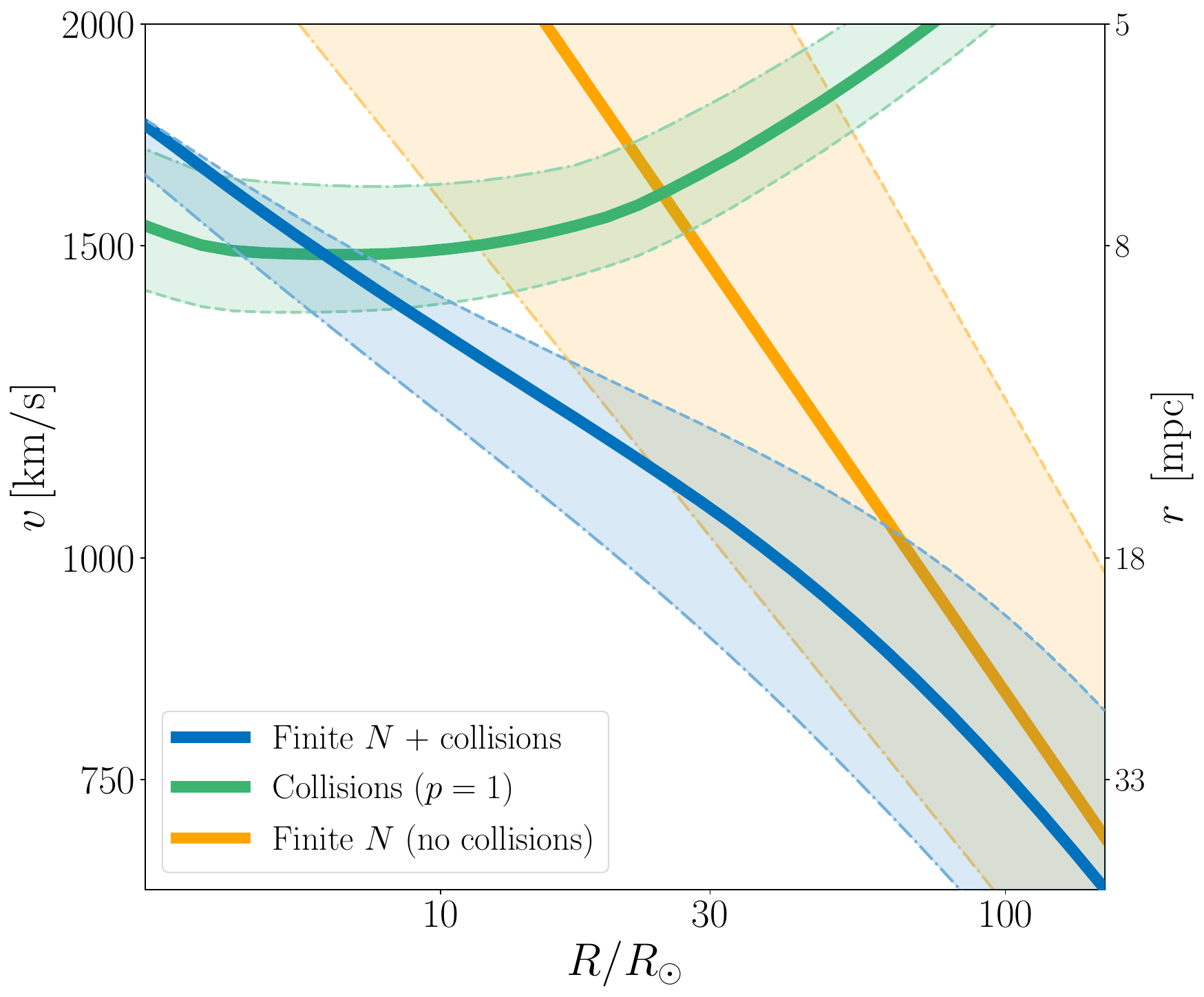}
    \caption{Characteristic velocities at which RGs are depleted as a function of their size.
    The green line shows the velocity where the collision probability over an RG lifetime reaches unity. 
    The yellow line gives the velocity where a single RG is expected to reside, assuming a power-law density profile (Eq.~\ref{eq:nrg}). 
    The blue line accounts for both the power-law distribution and the integrated RG survival probability (Eq.~\ref{eq:P}).
    For larger RGs, depletion mostly reflects their intrinsic scarcity, whereas for smaller RGs it is mainly set by collisions.
    The shaded regions correspond to a range of density slopes for the background stellar distribution, from $\gamma=5/4$ (dashed-dotted boundary) to $\gamma=7/4$ (dashed boundary), with $\gamma=3/2$ adopted as our fiducial value.
    }   
    \label{fig:r_dep}
\end{figure}

This phase-space structure translates into characteristic velocities (and radii), as shown in Fig.~(\ref{fig:r_dep}), above which RGs are expected to be depleted (blue line).
For comparison, we also present the corresponding velocities derived solely from the isotropic distribution and the RG short lifetimes (yellow line; Eq.~\ref{eq:aN1}) or from collisions (green line), defined by the radius where the collision timescale for circular orbits equals the RG lifetime (i.e., where $p=1$).

The impact of collisions can be somewhat disentangled from the intrinsic scarcity of RGs. 
From the scaling in Eq.~(\ref{eq:ps}), collisions deplete the RG population over a relatively narrow range in velocity.
Differentiating $p$ with respect to $R$ and $v$ yields
\begin{equation}
    \left.\frac{\dd v}{\dd R}\right|_{p={\rm const.}}\approx 0.3\frac{v}{R},
\end{equation}
Numerically, we find the prefactor to be $\approx 0.25$ for $0.5 \lesssim p \lesssim 1$ and $R\gtrsim 20\Rsun$.

By contrast, the isotropic RG distribution (Eq.~\ref{eq:aN1}) implies that larger RGs are absent at lower velocities (i.e., at greater distances from the SMBH). In this case,
\begin{equation}
    \left.\frac{\dd v}{\dd R}\right|_{\rm iso}\approx -0.5\frac{v}{R}.
\end{equation}
Therefore, comparing the characteristic distances at which RGs of different sizes begin to be depleted may help distinguish whether their absence is driven by collisions or simply reflects their overall intrinsic scarcity.

\begin{figure*}[ht!]
    \centering
    \includegraphics[width=\linewidth]{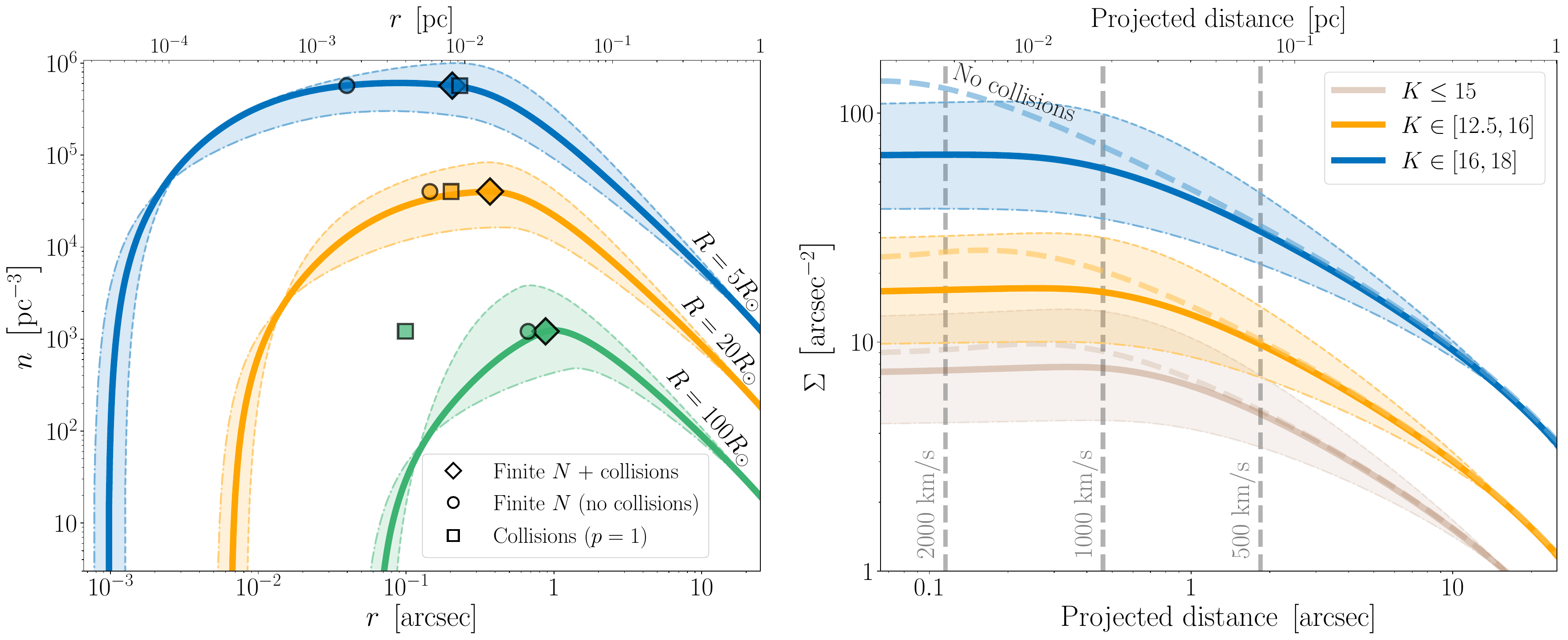}
    \caption{
    {\bf Left panel:} Number density of RGs with $R=5\Rsun$ (blue line), $R=20\Rsun$ (yellow line), and $R=100\Rsun$ (green line). In the outer regions ($r\gtrsim0.1\pc$), the RGs follow the underlying stellar density profile, taken as $\gamma=3/2$ (solid lines).
    The transition to a steeply declining profiles marks the depleted region. Diamonds markers denote the peak of the distribution, where circular orbits begin to be depleted. Circles indicate the expected break in the absence of collisions, as set from the isotropic distribution and the RG short lifetime. Squares show the depletion radius inferred by equating the collision timescale with the RG lifetime.
    {\bf Right panel}: Projected density of RGs in $K$-band magnitude bins: a bright bin ($K\in\l(12.5,16\r)$; yellow line) and a dim bin ($K\in\l(16,18\r)$; blue line), following \cite{Cano_2018,Habibi_2019}. 
    We also include the $K\leq15$ bin, as appeared in \cite{Schodel_2020}.
    We normalize the number of RGs in the bright bin at $r=1\pc$ according to observations \citep{Cano_2018}.
    The reference profiles, determined solely by the isotropic distribution assumption and the RG lifetimes, are shown as shaded dashed lines.
    In both panels, the shaded regions show the range obtained by varying the density slope from $\gamma=5/4$ (dashed-dotted lower bound) to $\gamma=7/4$ (dashed upper bound).
    }
    \label{fig:Proj_den}
\end{figure*}

This is further illustrated in the left panel of Fig.~(\ref{fig:Proj_den}), which shows the number density of RGs for different stellar radii. 
At large distances, $r\gtrsim0.1\pc$, RGs follow the same density slope as MS stars, as suggested by observations \citep{Schodel_2018} and imposed by construction in our model.
However, at smaller radii the RG density turns over and declines sharply.

The markers in the left panel of Fig.~(\ref{fig:Proj_den}) correspond to the characteristic depletion distances as defined above: 
diamond markers indicate the peak number density, the radius where the survival-weighted number of RGs reaches unity; 
circles denote the depletion radius expected from the isotropic distribution alone; and squares mark the radius at which RGs are expected to undergo a destructive collision within their lifetime.

For larger RGs, depletion mainly reflects their intrinsic scarcity, as illustrated by the proximity of the depletion radius and the isotropic expectation (diamond and circle markers) for $R = 100\Rsun$ RG (green line). 
In contrast, collisions deplete smaller RGs at significantly larger radii than predicted by the isotropic distribution. E.g., for $R=5\Rsun$ RG (blue line), the depletion radius lies close to the collision-driven radius (diamond and square markers) and is about an order of magnitude larger than the isotropic prediction (circle marker).

From the RG number density we calculate the projected density, shown in the right panel of Fig.~(\ref{fig:Proj_den}), divided into two $K$-band magnitude bins: $K\in\l(12.5,16\r)$, corresponding to $10.5\leq R/\Rsun< 60$, and $K\in\l(16,18\r)$, corresponding to $4\leq R/\Rsun<10.5$. 

The $K\in\l[12.5,16\r]$ bin exhibits a roughly flat projected profile within $\lesssim1$'', resulting in an average central surface density of $\Sigma\approx9-15~{\rm arcsec}^{-2}$. This range corresponds to varying the ambient stellar population density slope between $\gamma=5/4-3/2$.
This estimated surface density is consistent with the observationally inferred value \citep{Habibi_2019,Schodel_2020}. 
Neglecting collisions increases the central surface density by approximately $20\%$ (for the fiducial $\gamma=3/2$) and produces a flat profile within a smaller projected radii.

Nonetheless, focusing on the brightest RGs ($K\leq15$), we find an average projected surface density of $\Sigma\approx4-7~{\rm arcsec}^{-2}$, which is higher than \cite{Schodel_2020} observed value of $\Sigma\sim1~{\rm arcsec}^{-2}$.
Possible origins of this mild discrepancy are briefly discussed in \S\ref{sec:sum}.

For the fainter bin, $K\in\l[16,18\r]$, we find an average surface density of $\Sigma\approx 30-50~{\rm arcsec}^{-2}$ within $1$'', consistent with the values reported by \cite{Schodel_2020}. 

\section{Discussion \& Summary} \label{sec:sum}
We analytically study the properties of RG collisions in nuclear stellar clusters around SMBHs.
We show that at high velocities $\gtrsim 10^3\kms$, RGs are efficiently depleted by MS stars that strip their envelopes. 
At lower velocities, encounters with stellar-mass BHs dominate the destruction of larger RGs, although their impact on the overall RG population remains small, as shown in Fig. (\ref{fig:coll_prob}). 
Accretion feedback in the form of winds driven by super-Eddington accretion may enhance how efficiently stellar-mass BH collisions unbind RG envelopes (Eq. \ref{eq:E_acc}). 
Again, however, this is unlikely to have a large impact of the overall RG population relative to MS star collisions.  

The destruction of RGs in stellar collisions generates a unique population of isolated (non-binary) low-mass He WDs. 
We estimate a formation rate of $\sim500\Gyr^{-1}$ with a weak dependence on SMBH mass (Eq. \ref{eq:rate_hewd_int}). 
This population is a source of WD-SMBH transients in SMBHs with masses of $\sim 10^{5-6}~\Msun$ that are too massive to tidally disrupt more conventional C/O WDs formed by single star evolution. 
By contrast, it is not clear if standard binary channels for forming He WDs can operate efficiently in galactic nuclei where wide binaries are easily dynamically disrupted.

We find that RG envelope stripping is primarily determined by the impactor’s ability to deposit sufficient energy to unbind the envelope (Eq.~\ref{eq:E_ram}).
Subsequently, the generated shock can propagate inward, as the pressure only mildly increase throughout most of the RG envelope, enabling global mass loss of the full envelope in a single collision, as illustrated in Fig.~(\ref{fig:pressure}). 

In contrast, for MS star collisions, the internal pressure gradient imposes a stricter condition: even when enough energy is deposited to formally unbind the stars, a fully destructive collision requires the shock to overcome the pressure contrast between the center and the outer regions. 
For example, in a head-on collision of solar-type stars, we predict that the required velocity for the shock to unbind the core exceeds the escape velocity by roughly $\l(v/v_{{\rm esc},\odot}\r)^2\approx 20$, in excellent agreement with the numerical results of \cite{Rose_2026}. 

In the context of the GC, we estimate the number density of RGs and their corresponding projected density across different $K$-band magnitude bins, as shown in Fig. (\ref{fig:Proj_den}).
We normalize the number of RGs in the bright bin, $K\in\l(12.5,16\r)$, to observations at $1\pc$ \citep{Cano_2018}, where we further assume that they follow a similar density slope as the MS low-mass stars \citep[consistent with observational constraints, e.g.,][]{Schodel_2018}. 

Assuming that the RGs have an isotropic distribution, presumably inherited from their progenitor MS stars, and that the number of RGs with a given stellar radius scales with the time spent at that size, the most luminous (largest) RGs are unlikely to populate tightly bound or highly eccentric orbits simply due to their scarcity, as shown in Figs. (\ref{fig:tri}) and (\ref{fig:r_dep}).  
This leads to ``segregated'' depletion, in which larger RGs are absent at larger distances.

On top of that, collisions efficiently deplete the RG population at velocities above $\sim10^3\kms$, flattening the projected density profiles already at lower velocities, roughly at projected distances of $\sim1$'' (see Fig.~\ref{fig:Proj_den}). 
Unlike the absence arising from the isotropic distribution, collisions tend to deplete all RGs over a relatively narrow range of velocities.

Within the central $1$'', we find an average surface density of $\Sigma\approx9-15~{\rm arcsec}^{-2}$ for the $K\in\l[12.5,16\r]$ bin and $\gamma=5/4-3/2$. This is consistent with the observationally inferred value of $\Sigma\approx11{\rm arcsec}^{-2}$ \citep{Habibi_2019,Schodel_2020}. 
However, focusing only on the brightest RGs with $K\leq15$, our model predicts a surface density of $\Sigma\approx4-7~{\rm arcsec}^{-2}$, exceeding the $\Sigma\sim1~{\rm arcsec}^{-2}$ surface density inferred by \cite{Schodel_2020}. 
This discrepancy may reflect fluctuations in the GC stellar formation history, to which the brightest, short-lived RGs are particularly sensitive. 
It may also arise from small-number statistics or from our simplified assumption of power-law stellar distributions.

As noted above, the more recent observations \citep{Habibi_2019,Schodel_2020} indicate that the region from which RGs are ``missing'' is somewhat smaller and that the deficit is less severe than earlier work estimated.
Future observations, particularly those providing detailed spectro-photometric data, will place stronger constraints on stellar ages and enable a more precise characterization of the RG deficit, its radial extent, and its dependence on RG size. 
This, in turn, will help determine whether collisions alone are sufficient to resolve the missing RGs problem or if additional mechanisms, e.g., interactions with a clumpy disk \citep{Amaro_2020}, are required.

\begin{acknowledgments}
The authors would like to thank Jeremy Goodman for useful discussions. 
BR is supported by the Lyman Spitzer Jr. Fellowship.
\end{acknowledgments}

\appendix
\section{Multiple collisions} \label{app:b}
In \S\ref{sec:ram}, we determine the critical impact parameter $\bs$ required for a MS star to strip the entire RG envelope in a single encounter.
Collisions with larger impact parameters lead to a partial disruption of the envelope, removing only a fraction of its mass
\setcounter{equation}{0}
\begin{equation}
\frac{\delta m}{m_{\rm env}} \approx \left(\frac{v}{v_c}\right)^2,
\end{equation}
where $v_c$ is the velocity required to strip the entire envelope at a given impact parameter.

Consequently, the number of collisions required to fully strip the star is of order $N_{\rm col}\sim \l(v_c/v\r)^2$.
Thus, increasing the impact parameter raises the collision rate, but also increases the required number of collisions.
This trade-off restricts the optimal impact parameter to $b\sim R/2$. 
For grazing collisions with larger impact parameters, the impactor interacts only with the low-density outer layers; thus, the modest increase in collision rate from the larger cross section do not compensate for the substantially larger number of encounters needed to strip the envelope. 
Conversely, collisions with smaller impact parameter require fewer interactions but occur less frequently.

Therefore, the main enhancement from multiple encounters arises for MS-RG collisions, since for stellar-mass BHs the impact parameter required for full disruption in a single encounter is already comparable to the RG radius. 
\begin{figure}[h]
    \centering
    \includegraphics[width=8.6cm]{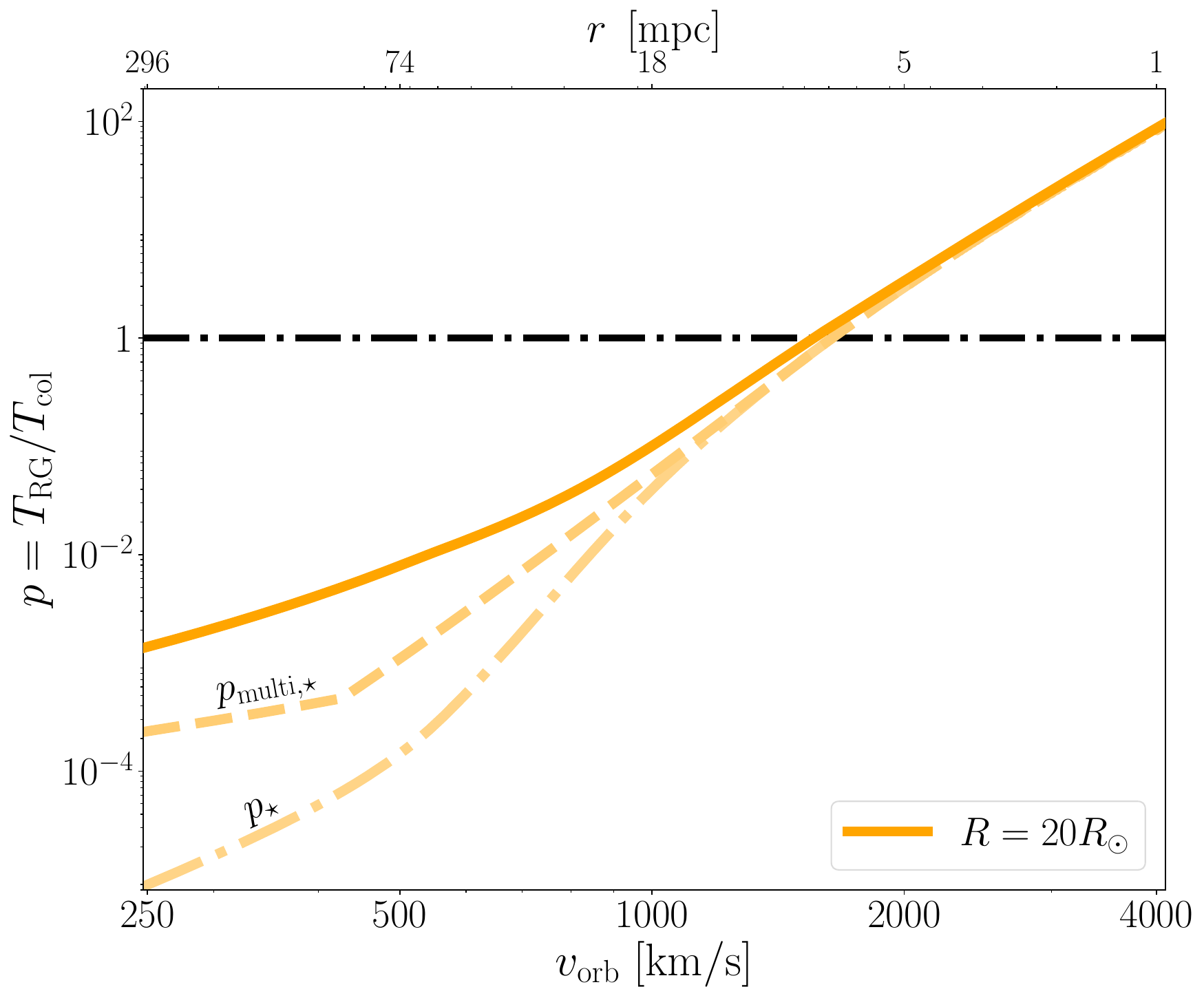}
    \caption{Collision probability as a function of orbital velocity for an $R=20\Rsun$ RG in a Milky Way-like galaxy. 
    The solid line shows the total probability for a single destructive collision (Eq. \ref{eq:p_i}), with stellar-mass BH collisions dominating at low velocity and MS-RG collisions dominating at higher velocity. The dashed line shows the contribution from multiple MS-RG collisions (Eq. \ref{eq:p_multi}), while the dash-dotted line corresponds to a single destructive MS-RG collision. 
    Although repeated collisions increase the MS–RG collision probability at low velocities, the overall disruption probability in this regime remains negligible and is dominated by encounters with stellar-mass BHs.
    The top x-axis indicates the distance from the SMBH.}
    \label{fig:rep_col}
\end{figure}

We define the probability that an RG is stripped via multiple collisions, analogously to Eq.~(\ref{eq:p_i}), as
\begin{equation}\label{eq:p_multi}
    p_{\rm multi,\star}=\frac{\trg/\Tc}{N_{\rm col}}.
\end{equation}
However, since for low velocities the single MS-RG collision probability is negligible, this enhancement does not suffice to significantly affect the RG population, as shown in Fig. (\ref{fig:rep_col}).

Finally, we assume that the RG structure is not significantly altered by partial envelope mass loss. 
This is motivated by the dominant role of the core in setting the equilibrium structure of the RG. Furthermore, the star can relax back to equilibrium on the Kelvin-Helmholtz timescale which is typically much shorter than the collision timescale ($T_{\rm KH} \lesssim 10^5~{\rm yr}$ for RGs with $R\gtrsim10\Rsun$). 

\section{Analytical approximation for the MS-RG collision}\label{app:a}
Beyond the order of magnitude estimate presented in Eqs.~(\ref{eq:rho_app})-(\ref{eq:ps}), we can obtain a more detailed analytic approximation for the maximal impact parameter required to deposit sufficient amount of energy into the envelope, as given by Eq. (\ref{eq:E_ram}).
Approximating the RG envelope density profile as
\begin{equation}
    \bar{\rho}\left(r\right)\approx \frac{\Msun}{2\pi R^{3}}\lrp{r}{R}{-3/2},
\end{equation}
we get
\begin{equation}\label{eq:surf_app}
\begin{aligned}
    \bar{\Sigma}_{\rm den}\left(b,R\right)& =2\int_b^R\frac{\bar\rho\left(r\right)r\dd r}{\left(r^2-b^2\right)^{1/2}} \approx 10^{11}{\rm g/cm^2}\lrp{R}{\Rsun}{-2} h\left(\frac{b}{R}\right),
\end{aligned}
\end{equation}
where 
\begin{equation}
    h(x)=\left(1-x^2\right)^{1/2}x^{-3/2}\widetilde{F}\left(1-\frac{1}{x^2}\right),
\end{equation}
$\widetilde{F}\l(x\r)={}_2F_1\left(\frac{1}{2},\frac{3}{4},\frac{3}{2},x\right)$, and ${}_2F_1$ is the hypergeometric function.

In the limit $b\ll R$, this simplifies to
\begin{equation}\label{eq:h_app}
    h(x\ll1)\approx 2\left(\frac{\pi^{1/2}\Gamma\left(5/4\right)}{\Gamma\left(3/4\right)}x^{-1/2}-1\right),
\end{equation}
where $\Gamma\left(x\right)$ is the Gamma function.

Using Eq.~(\ref{eq:E_ram}) together with Eqs.~(\ref{eq:surf_app}) and (\ref{eq:h_app}), we obtain:
 \begin{equation}\label{eq:b_app}
 \begin{aligned}
     \frac{\bar{b}_\star}{\Rsun}&\approx \frac{11.7Rv^4}{\l(2.6 v^2+1.4\times10^{5}R^{1.16}\r)^2},
 \end{aligned}
 \end{equation}
where R is given in units of $\Rsun$ and $v$ in $\rm km/s$.
In the above derivation, we consider an MS impactor and take its effective radius to be $\widetilde{R}_{\rm i}=\Rsun$, which is valid for $v\gtrsim v_{\rm esc,\odot}\sim600\kms$. At lower velocity, the Bondi-Hoyle radius exceeds the stellar radii.

As discussed in \S\ref{sec:ram}, the simplified power-law approximation $\rho\propto r^{-3/2}$ breaks down both in the inner regions of the RG, $R\lesssim0.5\Rsun$, near the He core and H-burning shell, as well as in the outer envelope, $R\gtrsim0.5R$, where the density drops steeply.
Hence, this approximation improves for larger RGs, for which it remains valid over a wider range of impact parameters and velocities.

\bibliography{main}{}

\begin{thebibliography}{}
\expandafter\ifx\csname natexlab\endcsname\relax\def\natexlab#1{#1}\fi
\providecommand{\url}[1]{\href{#1}{#1}}
\providecommand{\dodoi}[1]{doi:~\href{http://doi.org/#1}{\nolinkurl{#1}}}
\providecommand{\doeprint}[1]{\href{http://ascl.net/#1}{\nolinkurl{http://ascl.net/#1}}}
\providecommand{\doarXiv}[1]{\href{https://arxiv.org/abs/#1}{\nolinkurl{https://arxiv.org/abs/#1}}}

\bibitem[{{Aharon} \& {Perets}(2016)}]{Aharon_2016}
{Aharon}, D., \& {Perets}, H.~B. 2016, \apjl, 830, L1, \dodoi{10.3847/2041-8205/830/1/L1}

\bibitem[{Alexander(1999)}]{Alexander_1999}
Alexander, T. 1999, The Astrophysical Journal, 527, 835, \dodoi{10.1086/308129}

\bibitem[{{Alexander}(2017)}]{Alexander_2017}
{Alexander}, T. 2017, \araa, 55, 17, \dodoi{10.1146/annurev-astro-091916-055306}

\bibitem[{{Amaro Seoane}(2023)}]{Amaro_2023}
{Amaro Seoane}, P. 2023, \apj, 947, 8, \dodoi{10.3847/1538-4357/acb8b9}

\bibitem[{Amaro-Seoane \& Chen(2014)}]{Amaro_2014}
Amaro-Seoane, P., \& Chen, X. 2014, The Astrophysical Journal Letters, 781, L18, \dodoi{10.1088/2041-8205/781/1/L18}

\bibitem[{Amaro-Seoane {et~al.}(2020)Amaro-Seoane, Chen, Schödel, \& Casanellas}]{Amaro_2020}
Amaro-Seoane, P., Chen, X., Schödel, R., \& Casanellas, J. 2020, Monthly Notices of the Royal Astronomical Society, 492, 250, \dodoi{10.1093/mnras/stz3507}

\bibitem[{{Bahcall} \& {Wolf}(1976)}]{BW_76}
{Bahcall}, J.~N., \& {Wolf}, R.~A. 1976, \apj, 209, 214, \dodoi{10.1086/154711}

\bibitem[{{Bahcall} \& {Wolf}(1977)}]{BW_77}
---. 1977, \apj, 216, 883, \dodoi{10.1086/155534}

\bibitem[{Bailey \& Davies(1999)}]{Bailey_1999}
Bailey, V.~C., \& Davies, M.~B. 1999, Monthly Notices of the Royal Astronomical Society, 308, 257, \dodoi{10.1046/j.1365-8711.1999.02740.x}

\bibitem[{{Balberg}(2024)}]{Balberg_2024}
{Balberg}, S. 2024, \apj, 962, 150, \dodoi{10.3847/1538-4357/ad1690}

\bibitem[{{Balberg} {et~al.}(2013){Balberg}, {Sari}, \& {Loeb}}]{Balberg_2013}
{Balberg}, S., {Sari}, R., \& {Loeb}, A. 2013, \mnras, 434, L26, \dodoi{10.1093/mnrasl/slt071}

\bibitem[{{Bar-Or} \& {Alexander}(2016)}]{Bar_or_2016}
{Bar-Or}, B., \& {Alexander}, T. 2016, \apj, 820, 129, \dodoi{10.3847/0004-637X/820/2/129}

\bibitem[{{Bartko} {et~al.}(2010){Bartko}, {Martins}, {Trippe}, {Fritz}, {Genzel}, {Ott}, {Eisenhauer}, {Gillessen}, {Paumard}, {Alexander}, {Dodds-Eden}, {Gerhard}, {Levin}, {Mascetti}, {Nayakshin}, {Perets}, {Perrin}, {Pfuhl}, {Reid}, {Rouan}, {Zilka}, \& {Sternberg}}]{Bartko_2010}
{Bartko}, H., {Martins}, F., {Trippe}, S., {et~al.} 2010, \apj, 708, 834, \dodoi{10.1088/0004-637X/708/1/834}

\bibitem[{{Baumgardt} {et~al.}(2018){Baumgardt}, {Amaro-Seoane}, \& {Sch{\"o}del}}]{Baumgardt_2018}
{Baumgardt}, H., {Amaro-Seoane}, P., \& {Sch{\"o}del}, R. 2018, \aap, 609, A28, \dodoi{10.1051/0004-6361/201730462}

\bibitem[{{Binney} \& {Tremaine}(2008)}]{Biney_Tremaine}
{Binney}, J., \& {Tremaine}, S. 2008, {Galactic Dynamics: Second Edition}

\bibitem[{{Brown} {et~al.}(2020){Brown}, {Kilic}, {Kosakowski}, {Andrews}, {Heinke}, {Ag{\"u}eros}, {Camilo}, {Gianninas}, {Hermes}, \& {Kenyon}}]{ELM_Brown_2020}
{Brown}, W.~R., {Kilic}, M., {Kosakowski}, A., {et~al.} 2020, \apj, 889, 49, \dodoi{10.3847/1538-4357/ab63cd}

\bibitem[{{Buchholz} {et~al.}(2009){Buchholz}, {Sch{\"o}del}, \& {Eckart}}]{Buchholz_2009}
{Buchholz}, R.~M., {Sch{\"o}del}, R., \& {Eckart}, A. 2009, \aap, 499, 483, \dodoi{10.1051/0004-6361/200811497}

\bibitem[{{Cho} {et~al.}(2024){Cho}, {Prather}, {Su}, {Narayan}, \& {Natarajan}}]{Cho_2024}
{Cho}, H., {Prather}, B.~S., {Su}, K.-Y., {Narayan}, R., \& {Natarajan}, P. 2024, \apj, 977, 200, \dodoi{10.3847/1538-4357/ad9561}

\bibitem[{{Cohn} \& {Kulsrud}(1978)}]{Cohn_1978}
{Cohn}, H., \& {Kulsrud}, R.~M. 1978, \apj, 226, 1087, \dodoi{10.1086/156685}

\bibitem[{{Dale} {et~al.}(2009){Dale}, {Davies}, {Church}, \& {Freitag}}]{Dale_2009}
{Dale}, J.~E., {Davies}, M.~B., {Church}, R.~P., \& {Freitag}, M. 2009, \mnras, 393, 1016, \dodoi{10.1111/j.1365-2966.2008.14254.x}

\bibitem[{Do {et~al.}(2009)Do, Ghez, Morris, Lu, Matthews, Yelda, \& Larkin}]{Do_2009}
Do, T., Ghez, A.~M., Morris, M.~R., {et~al.} 2009, The Astrophysical Journal, 703, 1323, \dodoi{10.1088/0004-637X/703/2/1323}

\bibitem[{{Do} {et~al.}(2013){Do}, {Lu}, {Ghez}, {Morris}, {Yelda}, {Martinez}, {Wright}, \& {Matthews}}]{Do_2013}
{Do}, T., {Lu}, J.~R., {Ghez}, A.~M., {et~al.} 2013, \apj, 764, 154, \dodoi{10.1088/0004-637X/764/2/154}

\bibitem[{{Do} {et~al.}(2019){Do}, {Hees}, {Ghez}, {Martinez}, {Chu}, {Jia}, {Sakai}, {Lu}, {Gautam}, {O'Neil}, {Becklin}, {Morris}, {Matthews}, {Nishiyama}, {Campbell}, {Chappell}, {Chen}, {Ciurlo}, {Dehghanfar}, {Gallego-Cano}, {Kerzendorf}, {Lyke}, {Naoz}, {Saida}, {Sch{\"o}del}, {Takahashi}, {Takamori}, {Witzel}, \& {Wizinowich}}]{Do_2019}
{Do}, T., {Hees}, A., {Ghez}, A., {et~al.} 2019, Science, 365, 664, \dodoi{10.1126/science.aav8137}

\bibitem[{{Eisenhauer} {et~al.}(2005){Eisenhauer}, {Genzel}, {Alexander}, {Abuter}, {Paumard}, {Ott}, {Gilbert}, {Gillessen}, {Horrobin}, {Trippe}, {Bonnet}, {Dumas}, {Hubin}, {Kaufer}, {Kissler-Patig}, {Monnet}, {Str{\"o}bele}, {Szeifert}, {Eckart}, {Sch{\"o}del}, \& {Zucker}}]{Eisenhauer_2005}
{Eisenhauer}, F., {Genzel}, R., {Alexander}, T., {et~al.} 2005, \apj, 628, 246, \dodoi{10.1086/430667}

\bibitem[{{Event Horizon Telescope Collaboration} {et~al.}(2022){Event Horizon Telescope Collaboration}, {Akiyama}, {Alberdi}, {Alef}, {Algaba}, {Anantua}, {Asada}, {Azulay}, {Bach}, {Baczko}, {Ball}, {Balokovi{\'c}}, {Barrett}, {Baub{\"o}ck}, {Benson}, {Bintley}, {Blackburn}, {Blundell}, {Bouman}, {Bower}, {Boyce}, {Bremer}, {Brinkerink}, {Brissenden}, {Britzen}, {Broderick}, {Broguiere}, {Bronzwaer}, {Bustamante}, {Byun}, {Carlstrom}, {Ceccobello}, {Chael}, {Chan}, {Chatterjee}, {Chatterjee}, {Chen}, {Chen}, {Cheng}, {Cho}, {Christian}, {Conroy}, {Conway}, {Cordes}, {Crawford}, {Crew}, {Cruz-Osorio}, {Cui}, {Davelaar}, {De Laurentis}, {Deane}, {Dempsey}, {Desvignes}, {Dexter}, {Dhruv}, {Doeleman}, {Dougal}, {Dzib}, {Eatough}, {Emami}, {Falcke}, {Farah}, {Fish}, {Fomalont}, {Ford}, {Fraga-Encinas}, {Freeman}, {Friberg}, {Fromm}, {Fuentes}, {Galison}, {Gammie}, {Garc{\'\i}a}, {Gentaz}, {Georgiev}, {Goddi}, {Gold}, {G{\'o}mez-Ruiz}, {G{\'o}mez}, {Gu}, {Gurwell}, {Hada}, {Haggard}, {Haworth}, {Hecht}, {Hesper},
  {Heumann}, {Ho}, {Ho}, {Honma}, {Huang}, {Huang}, {Hughes}, {Ikeda}, {Impellizzeri}, {Inoue}, {Issaoun}, {James}, {Jannuzi}, {Janssen}, {Jeter}, {Jiang}, {Jim{\'e}nez-Rosales}, {Johnson}, {Jorstad}, {Joshi}, {Jung}, {Karami}, {Karuppusamy}, {Kawashima}, {Keating}, {Kettenis}, {Kim}, {Kim}, {Kim}, {Kim}, {Kino}, {Koay}, {Kocherlakota}, {Kofuji}, {Koch}, {Koyama}, {Kramer}, {Kramer}, {Krichbaum}, {Kuo}, {La Bella}, {Lauer}, {Lee}, {Lee}, {Leung}, {Levis}, {Li}, {Lico}, {Lindahl}, {Lindqvist}, {Lisakov}, {Liu}, {Liu}, {Liuzzo}, {Lo}, {Lobanov}, {Loinard}, {Lonsdale}, {Lu}, {Mao}, {Marchili}, {Markoff}, {Marrone}, {Marscher}, {Mart{\'\i}-Vidal}, {Matsushita}, {Matthews}, {Medeiros}, {Menten}, {Michalik}, {Mizuno}, {Mizuno}, {Moran}, {Moriyama}, {Moscibrodzka}, {M{\"u}ller}, {Mus}, {Musoke}, {Myserlis}, {Nadolski}, {Nagai}, {Nagar}, {Nakamura}, {Narayan}, {Narayanan}, {Natarajan}, {Nathanail}, {Fuentes}, {Neilsen}, {Neri}, {Ni}, {Noutsos}, {Nowak}, {Oh}, {Okino}, {Olivares}, {Ortiz-Le{\'o}n}, {Oyama},
  {{\"O}zel}, {Palumbo}, {Paraschos}, {Park}, {Parsons}, {Patel}, {Pen}, {Pesce}, {Pi{\'e}tu}, {Plambeck}, {PopStefanija}, {Porth}, {P{\"o}tzl}, {Prather}, {Preciado-L{\'o}pez}, \& {Psaltis}}]{EHT_SgrAI_2022}
{Event Horizon Telescope Collaboration}, {Akiyama}, K., {Alberdi}, A., {et~al.} 2022, \apjl, 930, L12, \dodoi{10.3847/2041-8213/ac6674}

\bibitem[{{Freitag} \& {Benz}(2002)}]{Freitag_2002}
{Freitag}, M., \& {Benz}, W. 2002, \aap, 394, 345, \dodoi{10.1051/0004-6361:20021142}

\bibitem[{{Freitag} \& {Benz}(2005)}]{Freitag_2005}
---. 2005, \mnras, 358, 1133, \dodoi{10.1111/j.1365-2966.2005.08770.x}

\bibitem[{{Gallego-Cano} {et~al.}(2018){Gallego-Cano}, {Sch{\"o}del}, {Dong}, {Nogueras-Lara}, {Gallego-Calvente}, {Amaro-Seoane}, \& {Baumgardt}}]{Cano_2018}
{Gallego-Cano}, E., {Sch{\"o}del}, R., {Dong}, H., {et~al.} 2018, \aap, 609, A26, \dodoi{10.1051/0004-6361/201730451}

\bibitem[{{Genzel} {et~al.}(2010){Genzel}, {Eisenhauer}, \& {Gillessen}}]{GenEisGil_2010}
{Genzel}, R., {Eisenhauer}, F., \& {Gillessen}, S. 2010, Reviews of Modern Physics, 82, 3121, \dodoi{10.1103/RevModPhys.82.3121}

\bibitem[{Genzel {et~al.}(1996)Genzel, Thatte, Krabbe, Kroker, \& Tacconi-Garman}]{Genzel_1996}
Genzel, R., Thatte, N., Krabbe, A., Kroker, H., \& Tacconi-Garman, L.~E. 1996, The Astrophysical Journal, 472, 153, \dodoi{10.1086/178051}

\bibitem[{{Genzel} {et~al.}(2003){Genzel}, {Sch{\"o}del}, {Ott}, {Eisenhauer}, {Hofmann}, {Lehnert}, {Eckart}, {Alexander}, {Sternberg}, {Lenzen}, {Cl{\'e}net}, {Lacombe}, {Rouan}, {Renzini}, \& {Tacconi-Garman}}]{Genzel_2003}
{Genzel}, R., {Sch{\"o}del}, R., {Ott}, T., {et~al.} 2003, \apj, 594, 812, \dodoi{10.1086/377127}

\bibitem[{{Ghez} {et~al.}(2003){Ghez}, {Duch{\^e}ne}, {Matthews}, {Hornstein}, {Tanner}, {Larkin}, {Morris}, {Becklin}, {Salim}, {Kremenek}, {Thompson}, {Soifer}, {Neugebauer}, \& {McLean}}]{GhezDuch_2003}
{Ghez}, A.~M., {Duch{\^e}ne}, G., {Matthews}, K., {et~al.} 2003, \apjl, 586, L127, \dodoi{10.1086/374804}

\bibitem[{{Ghez} {et~al.}(2008){Ghez}, {Salim}, {Weinberg}, {Lu}, {Do}, {Dunn}, {Matthews}, {Morris}, {Yelda}, {Becklin}, {Kremenek}, {Milosavljevic}, \& {Naiman}}]{Ghez_2008}
{Ghez}, A.~M., {Salim}, S., {Weinberg}, N.~N., {et~al.} 2008, \apj, 689, 1044, \dodoi{10.1086/592738}

\bibitem[{{Gibson} {et~al.}(2025){Gibson}, {Kiro{\u{g}}lu}, {Lombardi}, {Rose}, {Vanderzyden}, {Mockler}, {Gallegos-Garcia}, {Kremer}, {Ramirez-Ruiz}, \& {Rasio}}]{Gibson_2025}
{Gibson}, C. F.~A., {Kiro{\u{g}}lu}, F., {Lombardi}, J.~C., {et~al.} 2025, \apj, 980, 109, \dodoi{10.3847/1538-4357/ad9b80}

\bibitem[{{Gillessen} {et~al.}(2009){Gillessen}, {Eisenhauer}, {Trippe}, {Alexander}, {Genzel}, {Martins}, \& {Ott}}]{Gillessen_2009}
{Gillessen}, S., {Eisenhauer}, F., {Trippe}, S., {et~al.} 2009, \apj, 692, 1075, \dodoi{10.1088/0004-637X/692/2/1075}

\bibitem[{{GRAVITY Collaboration} {et~al.}(2018){GRAVITY Collaboration}, {Abuter}, {Amorim}, {Anugu}, {Baub{\"o}ck}, {Benisty}, {Berger}, {Blind}, {Bonnet}, {Brandner}, {Buron}, {Collin}, {Chapron}, {Cl{\'e}net}, {Coud{\'e} Du Foresto}, {de Zeeuw}, {Deen}, {Delplancke-Str{\"o}bele}, {Dembet}, {Dexter}, {Duvert}, {Eckart}, {Eisenhauer}, {Finger}, {F{\"o}rster Schreiber}, {F{\'e}dou}, {Garcia}, {Garcia Lopez}, {Gao}, {Gendron}, {Genzel}, {Gillessen}, {Gordo}, {Habibi}, {Haubois}, {Haug}, {Hau{\ss}mann}, {Henning}, {Hippler}, {Horrobin}, {Hubert}, {Hubin}, {Jimenez Rosales}, {Jochum}, {Jocou}, {Kaufer}, {Kellner}, {Kendrew}, {Kervella}, {Kok}, {Kulas}, {Lacour}, {Lapeyr{\`e}re}, {Lazareff}, {Le Bouquin}, {L{\'e}na}, {Lippa}, {Lenzen}, {M{\'e}rand}, {M{\"u}ler}, {Neumann}, {Ott}, {Palanca}, {Paumard}, {Pasquini}, {Perraut}, {Perrin}, {Pfuhl}, {Plewa}, {Rabien}, {Ram{\'\i}rez}, {Ramos}, {Rau}, {Rodr{\'\i}guez-Coira}, {Rohloff}, {Rousset}, {Sanchez-Bermudez}, {Scheithauer}, {Sch{\"o}ller}, {Schuler}, {Spyromilio},
  {Straub}, {Straubmeier}, {Sturm}, {Tacconi}, {Tristram}, {Vincent}, {von Fellenberg}, {Wank}, {Waisberg}, {Widmann}, {Wieprecht}, {Wiest}, {Wiezorrek}, {Woillez}, {Yazici}, {Ziegler}, \& {Zins}}]{Grav_col_2018}
{GRAVITY Collaboration}, {Abuter}, R., {Amorim}, A., {et~al.} 2018, \aap, 615, L15, \dodoi{10.1051/0004-6361/201833718}

\bibitem[{{GRAVITY Collaboration} {et~al.}(2020){GRAVITY Collaboration}, {Abuter}, {Amorim}, {Baub{\"o}ck}, {Berger}, {Bonnet}, {Brandner}, {Cardoso}, {Cl{\'e}net}, {de Zeeuw}, {Dexter}, {Eckart}, {Eisenhauer}, {F{\"o}rster Schreiber}, {Garcia}, {Gao}, {Gendron}, {Genzel}, {Gillessen}, {Habibi}, {Haubois}, {Henning}, {Hippler}, {Horrobin}, {Jim{\'e}nez-Rosales}, {Jochum}, {Jocou}, {Kaufer}, {Kervella}, {Lacour}, {Lapeyr{\`e}re}, {Le Bouquin}, {L{\'e}na}, {Nowak}, {Ott}, {Paumard}, {Perraut}, {Perrin}, {Pfuhl}, {Rodr{\'\i}guez-Coira}, {Shangguan}, {Scheithauer}, {Stadler}, {Straub}, {Straubmeier}, {Sturm}, {Tacconi}, {Vincent}, {von Fellenberg}, {Waisberg}, {Widmann}, {Wieprecht}, {Wiezorrek}, {Woillez}, {Yazici}, \& {Zins}}]{Grav_col_2020}
---. 2020, \aap, 636, L5, \dodoi{10.1051/0004-6361/202037813}

\bibitem[{{GRAVITY Collaboration} {et~al.}(2022){GRAVITY Collaboration}, {Abuter}, {Aimar}, {Amorim}, {Ball}, {Baub{\"o}ck}, {Berger}, {Bonnet}, {Bourdarot}, {Brandner}, {Cardoso}, {Cl{\'e}net}, {Dallilar}, {Davies}, {de Zeeuw}, {Dexter}, {Drescher}, {Eisenhauer}, {F{\"o}rster Schreiber}, {Foschi}, {Garcia}, {Gao}, {Gendron}, {Genzel}, {Gillessen}, {Habibi}, {Haubois}, {Hei{\ss}el}, {Henning}, {Hippler}, {Horrobin}, {Jochum}, {Jocou}, {Kaufer}, {Kervella}, {Lacour}, {Lapeyr{\`e}re}, {Le Bouquin}, {L{\'e}na}, {Lutz}, {Ott}, {Paumard}, {Perraut}, {Perrin}, {Pfuhl}, {Rabien}, {Shangguan}, {Shimizu}, {Scheithauer}, {Stadler}, {Stephens}, {Straub}, {Straubmeier}, {Sturm}, {Tacconi}, {Tristram}, {Vincent}, {von Fellenberg}, {Widmann}, {Wieprecht}, {Wiezorrek}, {Woillez}, {Yazici}, \& {Young}}]{Grav_col_2022b}
{GRAVITY Collaboration}, {Abuter}, R., {Aimar}, N., {et~al.} 2022, \aap, 657, L12, \dodoi{10.1051/0004-6361/202142465}

\bibitem[{{Guo} {et~al.}(2025){Guo}, {Stone}, {Quataert}, \& {Springel}}]{Guo_2025}
{Guo}, M., {Stone}, J.~M., {Quataert}, E., \& {Springel}, V. 2025, \apj, 987, 202, \dodoi{10.3847/1538-4357/add1da}

\bibitem[{{Habibi} {et~al.}(2019){Habibi}, {Gillessen}, {Pfuhl}, {Eisenhauer}, {Plewa}, {von Fellenberg}, {Widmann}, {Ott}, {Gao}, {Waisberg}, {Baub{\"o}ck}, {Jimenez-Rosales}, {Dexter}, {de Zeeuw}, \& {Genzel}}]{Habibi_2019}
{Habibi}, M., {Gillessen}, S., {Pfuhl}, O., {et~al.} 2019, \apjl, 872, L15, \dodoi{10.3847/2041-8213/ab03cf}

\bibitem[{{Hirai} {et~al.}(2018){Hirai}, {Podsiadlowski}, \& {Yamada}}]{Hirai_2018}
{Hirai}, R., {Podsiadlowski}, P., \& {Yamada}, S. 2018, \apj, 864, 119, \dodoi{10.3847/1538-4357/aad6a0}

\bibitem[{Jermyn {et~al.}(2023)Jermyn, Bauer, Schwab, Farmer, Ball, Bellinger, Dotter, Joyce, Marchant, Mombarg, Wolf, Sunny~Wong, Cinquegrana, Farrell, Smolec, Thoul, Cantiello, Herwig, Toloza, Bildsten, Townsend, \& Timmes}]{MESA_2023}
Jermyn, A.~S., Bauer, E.~B., Schwab, J., {et~al.} 2023, The Astrophysical Journal Supplement Series, 265, 15, \dodoi{10.3847/1538-4365/acae8d}

\bibitem[{Kim \& Goodman(2026)}]{Kim_2026}
Kim, T., \& Goodman, J. 2026, The Astrophysical Journal, 1001, 109, \dodoi{10.3847/1538-4357/ae4ed9}

\bibitem[{Kippenhahn {et~al.}(2012)Kippenhahn, Weigert, \& Weiss}]{Kippenhahn_2012}
Kippenhahn, R., Weigert, A., \& Weiss, A. 2012, {Stellar structure and evolution}, Astronomy and Astrophysics Library (Springer), \dodoi{10.1007/978-3-642-30304-3}

\bibitem[{Kormendy \& Ho(2013)}]{Kormendy_2013}
Kormendy, J., \& Ho, L.~C. 2013, Annual Review of Astronomy and Astrophysics, 51, 511, \dodoi{10.1146/annurev-astro-082708-101811}

\bibitem[{{Kremer} {et~al.}(2022){Kremer}, {Lombardi}, {Lu}, {Piro}, \& {Rasio}}]{Kremer_2022}
{Kremer}, K., {Lombardi}, J.~C., {Lu}, W., {Piro}, A.~L., \& {Rasio}, F.~A. 2022, \apj, 933, 203, \dodoi{10.3847/1538-4357/ac714f}

\bibitem[{{Lai} {et~al.}(1993){Lai}, {Rasio}, \& {Shapiro}}]{Lai_1993}
{Lai}, D., {Rasio}, F.~A., \& {Shapiro}, S.~L. 1993, \apj, 412, 593, \dodoi{10.1086/172946}

\bibitem[{{Law-Smith} {et~al.}(2017){Law-Smith}, {MacLeod}, {Guillochon}, {Macias}, \& {Ramirez-Ruiz}}]{Smith_2017}
{Law-Smith}, J., {MacLeod}, M., {Guillochon}, J., {Macias}, P., \& {Ramirez-Ruiz}, E. 2017, \apj, 841, 132, \dodoi{10.3847/1538-4357/aa6ffb}

\bibitem[{{Linial} \& {Metzger}(2023)}]{Linial_Metzger_2023}
{Linial}, I., \& {Metzger}, B.~D. 2023, \apj, 957, 34, \dodoi{10.3847/1538-4357/acf65b}

\bibitem[{{Linial} \& {Sari}(2022)}]{Linial_2022}
{Linial}, I., \& {Sari}, R. 2022, \apj, 940, 101, \dodoi{10.3847/1538-4357/ac9bfd}

\bibitem[{{Lu} \& {Quataert}(2023)}]{LuQua_2023}
{Lu}, W., \& {Quataert}, E. 2023, \mnras, 524, 6247, \dodoi{10.1093/mnras/stad2203}

\bibitem[{{MacLeod} {et~al.}(2013){MacLeod}, {Ramirez-Ruiz}, {Grady}, \& {Guillochon}}]{MacLeod_2013}
{MacLeod}, M., {Ramirez-Ruiz}, E., {Grady}, S., \& {Guillochon}, J. 2013, \apj, 777, 133, \dodoi{10.1088/0004-637X/777/2/133}

\bibitem[{{Magorrian} \& {Tremaine}(1999)}]{Magorrian_1999}
{Magorrian}, J., \& {Tremaine}, S. 1999, \mnras, 309, 447, \dodoi{10.1046/j.1365-8711.1999.02853.x}

\bibitem[{{Maguire} {et~al.}(2020){Maguire}, {Eracleous}, {Jonker}, {MacLeod}, \& {Rosswog}}]{Maguire_2020}
{Maguire}, K., {Eracleous}, M., {Jonker}, P.~G., {MacLeod}, M., \& {Rosswog}, S. 2020, \ssr, 216, 39, \dodoi{10.1007/s11214-020-00661-2}

\bibitem[{{Marietta} {et~al.}(2000){Marietta}, {Burrows}, \& {Fryxell}}]{Marietta_2000}
{Marietta}, E., {Burrows}, A., \& {Fryxell}, B. 2000, \apjs, 128, 615, \dodoi{10.1086/313392}

\bibitem[{{Merritt}(2004)}]{Merritt_2004}
{Merritt}, D. 2004, in Coevolution of Black Holes and Galaxies, ed. L.~C. {Ho}, 263, \dodoi{10.48550/arXiv.astro-ph/0301257}

\bibitem[{{Murphy} {et~al.}(1991){Murphy}, {Cohn}, \& {Durisen}}]{Murphy_1991}
{Murphy}, B.~W., {Cohn}, H.~N., \& {Durisen}, R.~H. 1991, \apj, 370, 60, \dodoi{10.1086/169793}

\bibitem[{Paxton {et~al.}(2010)Paxton, Bildsten, Dotter, Herwig, Lesaffre, \& Timmes}]{MESA_2011}
Paxton, B., Bildsten, L., Dotter, A., {et~al.} 2010, The Astrophysical Journal Supplement Series, 192, 3, \dodoi{10.1088/0067-0049/192/1/3}

\bibitem[{Paxton {et~al.}(2013)Paxton, Cantiello, Arras, Bildsten, Brown, Dotter, Mankovich, Montgomery, Stello, Timmes, \& Townsend}]{Mesa_2013}
Paxton, B., Cantiello, M., Arras, P., {et~al.} 2013, The Astrophysical Journal Supplement Series, 208, 4, \dodoi{10.1088/0067-0049/208/1/4}

\bibitem[{Paxton {et~al.}(2015)Paxton, Marchant, Schwab, Bauer, Bildsten, Cantiello, Dessart, Farmer, Hu, Langer, Townsend, Townsley, \& Timmes}]{MESA_2015}
Paxton, B., Marchant, P., Schwab, J., {et~al.} 2015, The Astrophysical Journal Supplement Series, 220, 15, \dodoi{10.1088/0067-0049/220/1/15}

\bibitem[{Paxton {et~al.}(2018)Paxton, Schwab, Bauer, Bildsten, Blinnikov, Duffell, Farmer, Goldberg, Marchant, Sorokina, Thoul, Townsend, \& Timmes}]{MESA_2018}
Paxton, B., Schwab, J., Bauer, E.~B., {et~al.} 2018, The Astrophysical Journal Supplement Series, 234, 34, \dodoi{10.3847/1538-4365/aaa5a8}

\bibitem[{Paxton {et~al.}(2019)Paxton, Smolec, Schwab, Gautschy, Bildsten, Cantiello, Dotter, Farmer, Goldberg, Jermyn, Kanbur, Marchant, Thoul, Townsend, Wolf, Zhang, \& Timmes}]{MESA_2019}
Paxton, B., Smolec, R., Schwab, J., {et~al.} 2019, The Astrophysical Journal Supplement Series, 243, 10, \dodoi{10.3847/1538-4365/ab2241}

\bibitem[{{Rauch}(1999)}]{Rauch_1999}
{Rauch}, K.~P. 1999, \apj, 514, 725, \dodoi{10.1086/306953}

\bibitem[{{Rees}(1988)}]{Rees_88}
{Rees}, M.~J. 1988, \nat, 333, 523, \dodoi{10.1038/333523a0}

\bibitem[{{Rizzuto} {et~al.}(2022){Rizzuto}, {Naab}, {Spurzem}, {Arca-Sedda}, {Giersz}, {Ostriker}, \& {Banerjee}}]{Rizzuto_2022}
{Rizzuto}, F.~P., {Naab}, T., {Spurzem}, R., {et~al.} 2022, \mnras, 512, 884, \dodoi{10.1093/mnras/stac231}

\bibitem[{{Rom} \& {Sari}(2025)}]{Rom_2025}
{Rom}, B., \& {Sari}, R. 2025, arXiv e-prints, arXiv:2502.13209, \dodoi{10.48550/arXiv.2502.13209}

\bibitem[{Rom \& Sari(2026)}]{Rom_2026}
Rom, B., \& Sari, R. 2026, \it in prep.

\bibitem[{{Rose} {et~al.}(2026){Rose}, {Lombardi}, {Gonz{\'a}lez Prieto}, {K{\i}ro{\u{g}}lu}, \& {Rasio}}]{Rose_2026}
{Rose}, S.~C., {Lombardi}, Jr., J.~C., {Gonz{\'a}lez Prieto}, E., {K{\i}ro{\u{g}}lu}, F., \& {Rasio}, F.~A. 2026, \apj, 1000, 162, \dodoi{10.3847/1538-4357/ae459b}

\bibitem[{Rose \& MacLeod(2024)}]{Rose_2024}
Rose, S.~C., \& MacLeod, M. 2024, The Astrophysical Journal, 963, L17, \dodoi{10.3847/2041-8213/ad251f}

\bibitem[{{Rose} {et~al.}(2022){Rose}, {Naoz}, {Sari}, \& {Linial}}]{Rose_2022}
{Rose}, S.~C., {Naoz}, S., {Sari}, R., \& {Linial}, I. 2022, \apjl, 929, L22, \dodoi{10.3847/2041-8213/ac6426}

\bibitem[{Rose {et~al.}(2023)Rose, Naoz, Sari, \& Linial}]{Rose_2023}
Rose, S.~C., Naoz, S., Sari, R., \& Linial, I. 2023, The Astrophysical Journal, 955, 30, \dodoi{10.3847/1538-4357/acee75}

\bibitem[{Ryu {et~al.}(2024)Ryu, Amaro Seoane, Taylor, \& Ohlmann}]{Ryu_2024}
Ryu, T., Amaro Seoane, P., Taylor, A.~M., \& Ohlmann, S.~T. 2024, Monthly Notices of the Royal Astronomical Society, 528, 6193, \dodoi{10.1093/mnras/stae396}

\bibitem[{{Sch{\"o}del} {et~al.}(2020){Sch{\"o}del}, {Nogueras-Lara}, {Gallego-Cano}, {Shahzamanian}, {Gallego-Calvente}, \& {Gardini}}]{Schodel_2020}
{Sch{\"o}del}, R., {Nogueras-Lara}, F., {Gallego-Cano}, E., {et~al.} 2020, \aap, 641, A102, \dodoi{10.1051/0004-6361/201936688}

\bibitem[{{Sch\"odel, R.} {et~al.}(2018){Sch\"odel, R.}, {Gallego-Cano, E.}, {Dong, H.}, {Nogueras-Lara, F.}, {Gallego-Calvente, A. T.}, {Amaro-Seoane, P.}, \& {Baumgardt, H.}}]{Schodel_2018}
{Sch\"odel, R.}, {Gallego-Cano, E.}, {Dong, H.}, {et~al.} 2018, A\&A, 609, A27, \dodoi{10.1051/0004-6361/201730452}

\bibitem[{Sellgren {et~al.}(1990)Sellgren, McGinn, Becklin, \& Hall}]{Sellgren_1990}
Sellgren, K., McGinn, M.~T., Becklin, E.~E., \& Hall, D.~B. 1990, The Astrophysical Journal, 359, 112.
\newblock \url{https://api.semanticscholar.org/CorpusID:122209017}

\bibitem[{{Syer} \& {Ulmer}(1999)}]{Syer_1999}
{Syer}, D., \& {Ulmer}, A. 1999, \mnras, 306, 35, \dodoi{10.1046/j.1365-8711.1999.02445.x}

\bibitem[{{Tuchman}(1985)}]{Tuchman_1985}
{Tuchman}, Y. 1985, \apj, 288, 248, \dodoi{10.1086/162785}

\bibitem[{{Wong} \& {Bildsten}(2025)}]{Wong_2025}
{Wong}, T. L.~S., \& {Bildsten}, L. 2025, \apj, 992, 108, \dodoi{10.3847/1538-4357/adfcd7}

\bibitem[{{Yao} {et~al.}(2025){Yao}, {Quataert}, {Jiang}, {Lu}, \& {White}}]{Yao_2025b}
{Yao}, P.~Z., {Quataert}, E., {Jiang}, Y.-F., {Lu}, W., \& {White}, C.~J. 2025, \apj, 978, 91, \dodoi{10.3847/1538-4357/ad8911}

\bibitem[{Zajaček {et~al.}(2020)Zajaček, Araudo, Karas, Czerny, \& Eckart}]{Zajaek_2020}
Zajaček, M., Araudo, A., Karas, V., Czerny, B., \& Eckart, A. 2020, The Astrophysical Journal, 903, 140, \dodoi{10.3847/1538-4357/abbd94}

\end{thebibliography}
\bibliographystyle{aasjournal}

\end{document}